\documentclass[%
 aip,
 amsmath,amssymb,
 reprint,%
]{revtex4-1}

\usepackage{graphicx}% Include figure files
\usepackage{dcolumn}% Align table columns on decimal point
\usepackage{bm}% bold math
\usepackage[utf8]{inputenc}
\usepackage[T1]{fontenc}
\usepackage{mathptmx}

\usepackage[capitalise]{cleveref}

\usepackage{tikz}
\usepackage{relsize}
\usepackage{diagbox}
\usepackage{tikz-cd} 

\usepackage{multirow}

\renewcommand{\epsilon}{\varepsilon}
\newcommand{\C}{\mathbb{C}}
\newcommand{\R}{\mathbb{R}}
\newcommand{\diag}{\textrm{diag}}
\newcommand{\rank}{\textrm{rank}}
\newcommand{\M}{\mathcal{M}}

\newcommand{\Her}{\textrm{Her}}

\newcommand{\ra}{\rangle}
\newcommand{\la}{\langle}
\newcommand{\mc}{\mathcal}

\newcommand{\tr}{\mathrm{tr}}

\newcommand{\nn}{\nonumber}

\newcommand{\psdrank}{\textrm{psd-rank}}
\newcommand{\purirank}{\textrm{puri-rank}}
\newcommand{\tipurirank}{\textrm{ti-puri-rank}}
\newcommand{\osr}{\textrm{osr}}
\newcommand{\hosr}{\textrm{hosr}}
\newcommand{\tiosr}{\textrm{ti-osr}}
\newcommand{\sqrtrank}{\textrm{sqrt-rank}}
\newcommand{\sr}{\textrm{SR}}
\newcommand{\qsqrtrank}{\textrm{q-sqrt-rank}}
\newcommand{\cpsdrank}{\textrm{cpsdt-rank}}
\newcommand{\cprank}{\textrm{cp-rank}}
\newcommand{\seprank}{\textrm{sep-rank}}
\newcommand{\tiseprank}{\textrm{ti-sep-rank}}
\newcommand{\symmrank}{\textrm{symm-rank}}

\newcommand{\be}{\begin{eqnarray}}
\newcommand{\ee}{\end{eqnarray}}
\newcommand{\ben}{\begin{enumerate}}
\newcommand{\een}{\end{enumerate}}
\newcommand{\bi}{\begin{itemize}}
\newcommand{\ei}{\end{itemize}}
\newcommand{\im}{\item}
\newcommand{\ba}{\begin{array}}
\newcommand{\ea}{\end{array}}

\newtheorem{theorem}{Theorem}

\newtheorem{corollary}[theorem]{Corollary}
\newtheorem{proposition}[theorem]{Proposition}

\newtheorem{definition}[theorem]{Definition}

\newtheorem{example}[theorem]{Example}
\newtheorem{remark}[theorem]{Remark}

\newcommand\xqed[1]{%
  \leavevmode\unskip\penalty9999 \hbox{}\nobreak\hfill
  \quad\hbox{#1}}
\newcommand\demo{\xqed{$\diamond$}}

\begin{document}

%\preprint{AIP/123-QED}

\title[Decompositions of mixed states in one spatial dimension]{Mixed states in one spatial dimension: \\ decompositions and correspondence with nonnegative matrices}

\author{Gemma de las Cuevas}
 \email{Gemma.DelasCuevas@uibk.ac.at}
 \affiliation{Institute for Theoretical Physics, Technikerstr.\ 21a,  A-6020 Innsbruck, Austria}

\author{Tim Netzer}%
 \email{Tim.Netzer@uibk.ac.at}
\affiliation{ 
Department of Mathematics, Technikerstr.\ 13,  A-6020 Innsbruck, Austria
}%

\date{\today}

\begin{abstract}
We study six natural decompositions of mixed states in one spatial dimension:  
the Matrix Product Density Operator (MPDO) form, 
the local purification form,
the separable decomposition (for separable states), 
and their three translational invariant (t.i.) analogues. 
For bipartite states diagonal in the computational basis, 
we show that these decompositions correspond to 
well-studied factorisations of an associated nonnegative matrix. 
Specifically, 
the first three decompositions correspond to the minimal factorisation, 
the nonnegative factorisation, 
and the positive semidefinite factorisation. 
We also show that a symmetric version of these decompositions corresponds to  
the symmetric factorisation, 
the completely positive factorisation, and
the completely positive semidefinite transposed factorisation, respectively. 
We leverage this correspondence to characterise  the six decompositions of mixed states.  
\end{abstract}

\maketitle

\section{Introduction}

Mixed states, namely positive semidefinite matrices of trace one, 
are not easy to characterise from a mathematical point of view. 
The reasons are at least threefold: 
first, they form a convex set with infinitely many extreme points, and therefore do not admit a concise description such as a vector space, or a convex set with ``corners''. 
Second, positive semidefinite matrices on a tensor product space $\mc{M}_d\otimes \mc{M}_d$ are not only given by convex combinations of positive semidefinite matrices on $\mc{M}_d$ and $\mc{M}_d$. 
And, third, the dimension of the vector space where they live grows exponentially with the number of subsystems, that is,  describing a state in $\mc{M}_d \otimes \cdots \otimes \mc{M}_d$ ($n$ times) requires roughly $ d^{2n}$ parameters. 
The third problem is not specific to positive semidefinite matrices, but is shared by  vectors $|\psi\ra \in \mathbb{C}^{d} \otimes \cdots \otimes \mathbb{C}^d$ (describing \emph{pure states}) and Hermitian operators (usually describing Hamiltonians) living in tensor product spaces.

The latter problem has  motivated the program of tensor networks, which aims at developing  efficient descriptions of quantum many-body systems \cite{Or18}. 
One of the central ideas of this program is that locally-based descriptions of states already capture many states of physical interest. 
While this has worked very well for pure states, 
mixed states seem to be more challenging, even in one spatial dimension. 
Some of the reasons for that are the difficulties associated to the description of positive semidefinite matrices mentioned above.

In this paper, we study local representations of  positive semidefinite (psd) matrices $\rho$ with a one-dimensional spatial structure. 
We aim to provide a comprehensive set of results regarding one-dimensional psd matrices; 
to this end, we will include known results and prove new ones. Our main tool to prove new results will be a correspondence between psd matrices and nonnegative matrices that we will establish in \cref{thm:corresp}; see below for further explanations.

Remark first that, physically, a one-dimensional structure means that $\rho$ describes the mixed state of a spin chain in one spatial dimension, so that, intuitively, the correlations between sites $i$ and $j$ are mediated by the sites inbetween, $i<l<j$. 
Mathematically, it means that $\rho$ is an element of a tensor product space with a natural order, 
$\mc{H}^{[1]} \otimes \mc{H}^{[2]} \otimes \mc{H}^{[3]}\otimes \cdots \otimes \mc{H}^{[n]}$, 
where $\mc{H}^{[l]}$ is the Hilbert space associated to site $l$, which is given by $\mc{M}_{d_l}$, the set of complex matrices of size $d_l\times d_l$. 
This natural order will be reflected in the fact that elements of $\mc{H}^{[l]}$ will share indices with $\mc{H}^{[l-1]}$  and $\mc{H}^{[l+1]}$  only.

%\red{Cite \cite{Gl19}}

We will analyse  the following decompositions of $\rho$: 
\begin{itemize}
\item[(i)] 
the Matrix Product Density Operator (MPDO) form, which is the most efficient representation, 
\item[(ii)]  
 the separable decomposition, which only exists for separable states, i.e.\ convex combination of positive semidefinite matrices on each tensor factor, and
\item[(iii)] 
 the local purification form, which has the advantage that the positivity is explicit in the local matrices,
 but which can be much more inefficient than the MPDO form \cite{De13c}, 
 \end{itemize} 

as well as their  translational invariant (t.i.) analogues: 
\begin{itemize}
\item[(iv)] the t.i.\ MPDO form, 
\item[(v)] the t.i.\ separable form  (for t.i.\ separable states) and
\item[(vi)] the t.i.\ local purification. 
\end{itemize}
For each decomposition we define a corresponding rank (see  \cref{tab:decomp}), which will be given by the minimum dimension of the tensors involved in that decomposition.

\begin{table}[htb]\centering
\begin{tabular}{cr|l }
&Decomposition of a psd matrix $\rho$& Minimal dimension\\ \hline
(i) &MPDO & operator Schmidt rank ($\osr$)\\
(ii) &separable decomposition & separable rank ($\seprank$)\\
(iii) &local purification & purification rank ($\purirank$)\\
(iv) &t.i.\ MPDO & t.i.\ operator Schmidt rank ($\tiosr$)\\
(v) &t.i.\ separable decomposition & t.i.\ separable rank ($\tiseprank$)\\
(vi) &t.i.\ local purification & t.i.\ purification rank ($\tipurirank$)
\end{tabular}
\caption{Decompositions for psd matrices considered in this paper, in the non-translational invariant and the translational invariant (t.i.) case, as well as their associated ranks.  } 
\label{tab:decomp}
\end{table}

For bipartite states which are diagonal in the computational basis, 
\be \label{eq:rhoM}
\rho =\sum_{i,j} m_{ij} |i,j\ra\la i,j|, 
\ee
we establish a correspondence between decompositions (i), (ii), (iii), and a symmetric version of (iv), (v) and (vi) of \cref{tab:decomp}, and factorisations of the nonnegative matrix\footnote{That is, entrywise nonnegative. A Hermitian matrix with nonnegative eigenvalues is called positive semidefinite.} 
\be \label{eq:M}
M = \sum_{i,j} m_{ij} |i\ra \la j|
\ee
presented in \cref{tab:nndecomp} (\cref{thm:corresp}). 
Most of these factorisations of nonnegative matrices have been defined previously, as well as their corresponding ranks (right column of \cref{tab:nndecomp}).

\begin{table}[htb]\centering 
\begin{tabular}{cr|l }
& Factorisation of a nonnegative matrix $M$ & Minimal dimension\\ \hline
(i) & minimal factorisation & rank\\
(ii) &nonnegative factorisation & nonnegative rank ($\rank_+$)\\
(iii) &positive semidefinite factorisation & psd rank ($\psdrank$)\\
(iv) &symmetric factorisation & symmetric rank ($\symmrank$)\\
(v) & completely positive factorisation  & cp rank ($\cprank$)\\
(vi) &completely psd transposed factorisation  & cpsdt rank ($\cpsdrank$)
\end{tabular}
\caption{Factorisations of a nonnegative matrix considered in \cref{thm:corresp}.}
\label{tab:nndecomp}
\end{table}

In words, we see the entries of a nonnegative matrix $M$ as the diagonal elements of a psd matrix of form \eqref{eq:rhoM}, and prove a correspondence of decompositions in   \cref{thm:corresp}. 
From our perspective, this is interesting for the study of decompositions of mixed states, as it provides a source of inspiration to generalise results (as illustrated in this paper, in \cite{De13c} and \cite{De19c}), 
as well as for the study of ranks of nonnegative matrices, 
as it provides a natural path to generalisation to the case that $\rho$ is not diagonal, and/or $\rho$ is multipartite.

In this paper, we will  illustrate the use of  \cref{thm:corresp} for decompositions of mixed states. 
Specifically,  in \cref{sec:bounds} and  \cref{sec:tibounds}, we will analyse and characterise the decompositions of psd matrices mentioned in \cref{tab:decomp} by proving several bounds and relations among the various ranks. 
 Many of the results will be generalisations of the corresponding results for nonnegative matrices, and some will be derived independently. 
We will point to more possible generalisations  in the outlook (\cref{sec:concl}).

The paper is meant to be accessible for researchers in quantum information and convex algebraic geometry, as we will rederive basic definitions.  
It is organized as follows:
\bi
\im In \cref{sec:pos} we present and analyse the MPDO form, 
the separable decomposition, and the local purification form. 
\im In \cref{sec:posti} we define and analyse the t.i.\ MPDO form, 
the t.i.\ separable decomposition, and the t.i.\ local purification.
\im In \cref{sec:matrixcones} we present factorisations of nonnegative matrices and   their correspondence with decompositions of psd matrices (\cref{thm:corresp}). 
\im In \cref{sec:bounds} we provide bounds for the non-t.i.\ decompositions.
\im In \cref{sec:tibounds} we provide bounds for the t.i.\ decompositions. 
\im In \cref{sec:concl} we  conclude and present an outlook. 
\im Finally, in  \cref{app:proofmain} we prove \cref{thm:corresp}.
\ei

%% ==========================================
\section{Decompositions of psd matrices}
\label{sec:pos}

In this section we present and analyse the relevant decompositions  and ranks for psd matrices 
in one spatial dimension in the non-t.i.\ case. 
First we present some general notions (\cref{ssec:gen}), 
then the Matrix Product Density Operator (MPDO) form (\cref{ssec:mpdo}),
the separable decomposition (\cref{ssec:sepdecomp}),
and finally the local purification form (\cref{ssec:lp}).

%%===========================
\subsection{General notions}
\label{ssec:gen}

Throughout this paper, 
a positive semidefinite  (psd) matrix  is a Hermitian matrix with nonnegative eigenvalues, and 
$\rho\geqslant 0$ denotes that $\rho$ is psd. 
Our main object of study is a psd matrix $\rho$ defined on an $n$-fold tensor product space, 
$$
0 \leqslant \rho \in \mc{H}^{[1]} \otimes \cdots \otimes \mc{H}^{[n]}, 
$$
where $\mc{H}^{[l]}$ is the Hilbert space associated to site $l$, 
which is identified with the space of complex matrices of size $d_l\times d_l$, denoted $\M_{d_l}$.\footnote{Everything is finite-dimensional in our discussion.} 
In some cases, for simplicity, we will assume that $d_l=d$ for all $l$. 
The Hilbert space $\mc{H}^{[1]} \otimes \cdots \otimes \mc{H}^{[n]}$  is often called the physical space, and its dimension $d_1^2\cdots d_n^2$ is  called the  physical dimension. 

\begin{remark}[Ignoring normalisation]\label{rem:norm}
In this paper we will ignore normalisation conditions on $\rho$ that are often imposed in physics, since considering  $\rho/\tr(\rho)$ instead of $\rho$ amounts to multiplying $\rho$  by a positive number, which does not change any of the ranks that we will analyse.  
For this reason, in this paper we will analyse decompositions of psd matrices, rather than decompositions of states. 
\demo\end{remark}

Let us recall some basic definitions. 

\begin{definition}\label{def:basic}
A psd matrix $0\leqslant \rho \in \mc{M}_{d_1} \otimes \cdots \otimes \mc{M}_{d_n} $ is 
\bi
\im  \emph{product} if it can be written as $\rho = A^{[1]} \otimes A^{[2]} \otimes \cdots \otimes A^{[n]}$ with $A^{[j]}\in \mc{M}_{d_j}$.
\im \emph{separable} if it can be written as a sum of product psd matrices. 
\im  \emph{entangled} if it is not separable. 
\im  \emph{pure} if $\rank(\rho)=1$.
\ei
\end{definition}

%%===============================
\subsection{The Matrix Product Density Operator form}
\label{ssec:mpdo}

 This subsection is devoted to the first natural way of representing a psd matrix $\rho$: the MPDO form. 
 The results of this section are not new; they can be found, or can be easily derived from, e.g., Refs.\ \cite{Ve04d,Zw04}.

\begin{definition}[MPDO]\label{def:mpdo}
Let  
$0\leqslant \rho \in \mc{H}^{[1]}\otimes \cdots \otimes  \mc{H}^{[n]}$. 
A \emph{Matrix Product Density Operator} (MPDO) form \cite{Ve04d} of $\rho$ is given by 
\be
\rho = \sum_{\alpha_1,\ldots,\alpha_{n-1}=1}^D 
A^{[1]}_{\alpha_1} \otimes 
A^{[2]}_{\alpha_1,\alpha_2} \otimes 
\cdots \otimes
A^{[n-1]}_{\alpha_{n-2},\alpha_{n-1}} \otimes 
A^{[n]}_{\alpha_{n-1}} ,
\label{eq:mpdo}
\ee
where $A^{[l]}_{\alpha}\in\mc{H}^{[l]}$ for $l=1,n$, and
$A^{[l]}_{\alpha,\alpha'}\in\mc{H}^{[l]}$ for $1<l<n$. 
The minimum such $D$ is called the operator Schmidt rank of $\rho$, denoted $\osr(\rho)$.
\end{definition}

Note that we are referring to it as {\it an} MPDO form instead of {\it the} MPDO form, because it is not unique
(see \cref{rem:freedom}).
To keep the notation simple, we will assume that each $\mc{H}^{[l]}$ is given by $\mc{M}_d$ throughout this section.

\begin{remark}[Expressing the physical indices]\label{rem:phys}
 In \cref{def:mpdo}, for $1<l<n$, we see the tensor $A^{[l]}$ as a collection 
 $(A_{\alpha,\beta}^{[l]}\in \M_d)_{\alpha,\beta=1}^D$. 
 We can also see $A^{[l]}$  as a collection $((A^{[l]})^{i,j}\in \M_D)_{i,j=1}^d$, where  Latin letters such as $i,j$ denote physical indices, and Greek letters such as $\alpha,\beta$ denote virtual indices. 
Similarly, $A^{[1]}$ and $A^{[n]}$ only have three indices, namely $i,j,\alpha$. 
While in \cref{def:mpdo} we see $A^{[1]}$ as a collection $\{A^{[1]}_\alpha \in \mc{M}_d\}_{\alpha=1}^D$,
 we can also see it as a set of row vectors $\{(A^{[1]})^{i,j} \in \mathbb{C}^D\}_{i,j=1}^d$. 
The situation is similar for $A^{[n]}$, with the only difference that $\{(A^{[n]})^{i,j}\}$ are column vectors. 
This allows us to write the physical indices explicitly  in  \eqref{eq:mpdo}, resulting in 
\be
\rho = \sum_{i_1,\ldots,i_n,j_1,\ldots,j_n=1}^d
(A^{[1]})^{i_1,j_1}(A^{[2]})^{i_2,j_2}\cdots (A^{[n]})^{i_n,j_n}
 \nonumber \\
|i_1,\ldots, i_n\ra\la j_1,\ldots,j_n| .
\nonumber
\ee 
\demo \end{remark}

\begin{remark}[Connection to tensor rank] \label{rem:trtompdo}
For $n=2$ the operator Schmidt rank is just the tensor rank, ${\rm tsr}(\rho)$, i.e.\ the minimal number of elementary tensors needed to obtain $\rho$ as their sum, since  an MPDO form is  just $$\rho=\sum_{\alpha=1}^D A^{[1]}_{\alpha}\otimes A^{[2]}_{\alpha}.$$
In general we  have
$$\osr(\rho)\leq {\rm tsr}(\rho)\leq\osr(\rho)^{n-1}.$$ 
The first inequality is obtained by starting with a decomposition $$\rho=\sum_{i=1}^{{\rm tsr}(\rho)} B_i^{[1]}\otimes B_i^{[2]}\otimes\cdots\otimes B_i^{[n]}$$ and defining $A_{\alpha}^{[l]}=B_\alpha^{[l]}$ for $l=1,n$, and 
$A_{\alpha,\beta}^{[l]}=\delta_{\alpha,\beta} B_{\alpha}^{[l]}$ for $l=2,\ldots, n-1.$ 
The second inequality is clear by counting the number of summands in  (\ref{eq:mpdo}).
\demo\end{remark}

\begin{remark}[Computing the MPDO form] \label{rem:constructionMPDO}
An MPDO form that realizes ${\rm osr}(\rho)$ can be obtained by doing successive singular value decompositions (SVD) between the  linear bipartitions $[1,\ldots, l] |[l+1, \ldots, n]$ for $1\leq l<n$ of $\rho$. Namely, first we do an SVD across bipartition $[1]|[2\ldots n]$, to obtain
\be\label{eqn:svd1}
\rho= 
\sum_{\alpha_1=1}^{D_1} A^{[1]}_{\alpha_1} \otimes E^{[2\ldots n]}_{\alpha_1},
\label{eq:rho-OSR1}
\ee
where $A^{[1]}_{\alpha} \in \mc{H}^{[1]}$, 
and $E^{[2\ldots n]}_{\alpha} \in \mc{H}^{[2]}\otimes \cdots \otimes \mc{H}^{[n]}$, where we have absorbed the singular values, say, in $A^{[1]}_{\alpha_1}$. 
$D_1$ is thus the rank of $\rho$ across  bipartition $[1]|[2,\ldots, n]$, and we have that $D_1\leq d^2$.

Now we consider the tensor $E^{[2\ldots n]} = \{E^{[2\ldots n]}_{\alpha_1}\}_{\alpha_1}$. 
Index $\alpha_1$ is associated to the Hilbert space of the first virtual system, denoted $\mc{H}^{[1^{\textrm{v}}]} = \mathbb{C}^{D_1}$, and therefore $E^{[2\ldots n]}$ is an element of  $\mc{H}^{[1^{\textrm{v}}]}  \otimes \mc{H}^{[2]}\otimes  \cdots \otimes\mc{H}^{[n]}$.
We now do an SVD across  bipartition $[1^{\textrm{v}},2]|[3,\ldots,n]$, to obtain
\be
\label{eq:D2}
E^{[2\ldots n]} = 
\sum_{\alpha_2=1}^{ D_2}
A^{[2]}_{\alpha_2} \otimes 
E^{[3\ldots n]}_{\alpha_2} ,
\ee
where $A^{[2]}_{\alpha} \in \mc{H}^{[1^{\textrm{v}}]}  \otimes \mc{H}^{[2]}$ and 
$E^{[3\ldots n]}_{\alpha} \in \mc{H}^{[3]}\otimes\ldots \otimes \mc{H}^{[n]}$. 
Writing down explicitly an element of $ \mc{H}^{[1^{\textrm{v}}]} $,  labeled by index $\alpha_1$, we obtain
$$
E^{[2\ldots n]}_{\alpha_1} = 
\sum_{\alpha_2=1}^{ D_2}
A^{[2]}_{\alpha_1,\alpha_2} \otimes 
E^{[3\ldots n]}_{\alpha_2} .
$$
Note that $D_2\leq D_1d^2$. Then we proceed similarly with the next tensor, $E^{[3\ldots n]}$, until  we obtain form \eqref{eq:mpdo}. Since for subsequent decompositions we will have that $D_l\leq D_{l-1}d^2$,  in general $D_l$ grows exponentially with $l$.

It is quite easy to see that  $D_l$ is the rank of $\rho$ across the bipartition $[1,\ldots, l] |[l+1, \ldots, n]$. 
This already shows that $D:=\max_l D_l$ is the minimum number such that a decomposition of the form \eqref{eq:mpdo} is possible.
\demo\end{remark}

\begin{remark}[Freedom in the decomposition]\label{rem:freedom}
To construct the MPDO form we do not really need the SVD, but any decomposition whose intermediate dimension is the rank. 
Namely, given a matrix $A \in \mathbb{C}^{p\times q}$ any decomposition $A=BC$ where $B$ has $r=\rank(A)$ columns works as well. 
In this paper we do not fix the freedom in the choice of $B,C$.

The analogue of the MPDO form for vectors $|\psi\ra \in \mathbb{C}^{d^n}$ gives rise to vectors in Matrix Product State form, or simply Matrix Product States  \cite{Pe07,Fa92}.  
This freedom is well characterised in this case, where it is fixed by choosing the so-called canonical form \cite{Pe07}, or its generalisation, the irreducible form \cite{De17}. 
\demo \end{remark}

\begin{remark}[The Hermitian MPDO form]
In  \cite{De19c}, a \emph{Hermitian} MPDO is introduced, which only differs from the MPDO form in the fact that the tensors $A^{[l]}_{\alpha,\beta}$ need to be Hermitian. The associated  minimal number of terms is called the Hermitian operator Schmidt rank, denoted $\hosr$. In Ref.\ \cite{De19c} it is shown that if $\rho$ is a bipartite psd matrix (i.e.\ $n=2$), then $\osr(\rho)=\hosr(\rho)$, but in the multipartite case, $\osr(\rho)\leq \hosr(\rho)\leq 2^{n-1}\osr(\rho)$, although we do not know whether the latter inequality is tight. 
In this paper we will refrain from analysing the hermitian operator Schmidt rank further.

We remark that one can force the local tensors to be  ``nearly Hermitian" by only doubling the number of terms $D$.
To see this, 
define a tensor $B^{[l]}\in \mc{M}_d ( \mc{M}_{2D})$ as
\begin{align*}
&(B^{[l]})^{i,j} = 2^{-1/n}\sum_{\alpha,\beta=1}^D \left(
(A^{[l]})^{i,j}_{\alpha,\beta} |\alpha\ra\la \beta| +  
 (\bar A^{[l]})^{j,i}_{\alpha,\beta} |\alpha+D\ra\la \beta+D| \right), \quad 1<l<n
\\ 
&
(B^{[1]})^{i,j} = 2^{-1/n}\sum_{\alpha=1}^D \left(
(A^{[1]})^{i,j}_{\alpha} \la \alpha| +  
 (\bar A^{[1]})^{j,i}_{\alpha} \la \alpha|\right),
\end{align*}
where $\bar{}$ denotes complex conjugate, and 
where the tensors $A^{[l]}$'s are those of \eqref{eq:mpdo}. $B^{[n]}$ is defined in the same way, and since $\rho$ is Hermitian, we have that $$
\rho = \sum_{i_1,\ldots,i_n, j_1,\ldots, j_n=1}^d\tr\left((B^{[1]})^{i_1,j_1}
\cdots (B^{[n]})^{i_n,j_n}\right) |i_1,\ldots,i_n\ra \la j_1,\ldots, j_n| , 
$$ 
and $B$ is ``nearly Hermitian" with respect to the physical indices $i,j$, as we need an additional permutation of the virtual indices: 
$$
B^{i,j}_{\alpha,\beta } = \bar B^{j,i}_{\beta\oplus D,\alpha\oplus D} ,
$$ 
 where $\oplus$ means sum modulo $2D$. 
\demo\end{remark}

%%------------------positivity ------------
The disadvantage of the MPDO form is that the local tensors $A^{[l]}_{\alpha,\alpha'}$ are not psd.
This is a challenge for the theoretical program of tensor networks, which aims at characterising the properties of $\rho$ in terms of the local tensors, such as the exponential decay of correlations \cite{Pe07,Sc10b}, symmetries \cite{Sc10b} or the existence of a continuum limit \cite{De17b}, to cite some examples.  
Yet, from the local tensors of the MPDO form one cannot characterise the most basic property of $\rho$, namely that it is psd. 
It is also problematic numerically, as a truncation of the auxiliary index (i.e.\ a replacement of $D$ by $\tilde D < D$) will generally destroy the positivity of $\rho$. 
Enforcing positivity in the local matrices leads to the local purification, which we discuss in \cref{ssec:lp}. 

%%-----------------------------
\begin{remark}[The Matrix Product Operator (MPO) form] \label{rem:mpo} 
Consider an operator 
\be
L\in \mc{M}_{d_1,d_1'}\otimes \cdots \otimes \mc{M}_{d_n,d_n'}, 
\label{eq:Lmporem}
\ee
which need not be a square matrix, and where $\mc{M}_{d,d'}$ denotes the space of complex matrices of size $d\times d'$. 
By the same construction as for the MPDO (see \cref{rem:constructionMPDO}), we can reach the so-called \emph{Matrix Product Operator} (MPO) form
\be
L = \sum_{\alpha_1,\ldots,\alpha_{n-1}=1}^{D}
C^{[1]}_{\alpha_1}\otimes
C^{[2]}_{\alpha_1,\alpha_2}\otimes\cdots \otimes
C^{[n]}_{\alpha_{n-1}} ,
\label{eq:Lpurif}
\ee
where $C^{[1]}_{\alpha} \in \mc{M}_{d_1,d_1'}$, 
$C^{[n]}_{\alpha} \in \mc{M}_{d_n,d_n'}$
and
$C^{[l]} \in\mc{M}_{d_l,d_l'}$ for all intermediate $l$'s.
The minimum such $D$ is also called the operator Schmidt rank of $L$, denoted $\osr(L)$. 
Indeed, the only difference between the MPO and the MPDO form is that in the latter the operator is globally psd. 
This highlights the fact that the construction of the MPDO form does not use that $\rho$ is psd. 

If $L$ is a vector (i.e.\ a column matrix), the operator Schmidt rank is called the Schmidt rank, usually. In this paper, nonetheless, we will still refer to it as the operator Schmidt rank of $L$ to avoid introducing new terminology. The same will be true for the t.i.\ operator Schmidt rank, to be introduced in \cref{def:tiMPDO}. 
\demo
\end{remark}

We conclude this section with some basic inequalities fulfilled by the operator Schmidt rank. 

\begin{proposition}\label{prop:osrineq}
Let  $\rho, \tau \in \mc{H}^{[1]}\otimes \cdots \otimes  \mc{H}^{[n]}$. Then
\begin{itemize}
\item[(i)] $\osr(\rho+\tau)\leq \osr(\rho)+\osr(\tau).$
\item[(ii)] $\osr(\rho\tau)\leq \osr(\rho)\osr(\tau).$
%\item[(iii)] $\osr(\rho\otimes\tau)=\max\{\osr(\rho),\osr(\tau)\}.$
\end{itemize}
\end{proposition}
%\begin{proof} 
\emph{Proof.}
If $A^{[i]}$ and $B^{[i]}$ are tensors that provide an MPO form for $\rho$ and $\tau$, respectively, then the block-diagonal sums $A^{[i]}\oplus B^{[i]}$ provide an MPO form for $\rho+\tau$, and the tensors $A^{[i]}\otimes B^{[i]}$ provide a MPO form for $\rho\tau$. 
\demo
%\end{proof}

%%==============================================
\subsection{The separable decomposition}
\label{ssec:sepdecomp}

While the MPDO form (and the later to be defined) local purification form exist for any psd matrix $\rho$, we now consider the separable decomposition, which  exists  only for separable psd matrices. 
 To the best of our knowledge, the separable decomposition and the associated separable rank are introduced here, although it is a very natural definition that may have been considered before.
We start by recalling the definition of separable psd matrix (\cref{def:basic}). 

\begin{definition}[Separable psd matrix]\label{def:sep} 
Let $0\leqslant \rho \in \mc{H}^{[1]} \otimes \cdots  \otimes \mc{H}^{[n]} $. 
We say that $\rho$ is \emph{separable} if it can be written as 
\be
\rho = \sum_{\alpha}  \: A^{[1]}_\alpha \otimes A^{[2]}_\alpha \otimes \cdots \otimes A^{[n]}_\alpha,
\label{eq:sepdef}
\ee
where
\be
A^{[1]}_\alpha \otimes A^{[2]}_\alpha \otimes \cdots \otimes A^{[n]}_\alpha  \geqslant 0
\label{eq:prod}
\ee
for all $\alpha$. 
\end{definition}

Note that separability is often defined in terms of convex combinations of psd product states. Since we are ignoring the normalisation  (see \cref{rem:norm}), we can consider sums instead of convex combinations. 
Note also that the condition on product psd matrices [Eq.\ \eqref{eq:prod}] implies that each $A_{\alpha}^{[l]}$ (for each $\alpha,l$) is semidefinite, that is, either psd or negative semidefinite, and that an even number of them is negative semidefinite. By redefining the negative semidefinite ones as $-A_{\alpha}^{[l]} $, we can assume w.l.o.g. that each $A_{\alpha}^{[l]}$ is psd.

\begin{definition}[Separable decomposition]\label{def:sepdecomp}
Let $0\leqslant \rho \in \mc{H}^{[1]} \otimes \cdots \otimes \mc{H}^{[n]} $ be separable. 
A \emph{separable decomposition} of  $\rho$ is given by 
\be
\rho = \sum_{\alpha_1,\ldots, \alpha_{n-1}=1}^D \chi^{[1]}_{\alpha_{1}} \otimes
\chi^{[2]}_{\alpha_{1},\alpha_2}  \otimes \cdots \otimes 
\chi^{[n]}_{\alpha_{n-1}}  
\label{eq:sep}
\ee
where each of these matrices is psd, i.e.\ $\chi^{[1]}_{\alpha}\geqslant 0$ and $\chi^{[n]}_{\alpha}\geqslant 0$,
  and $\chi^{[l]}_{\alpha,\beta}\geqslant 0 $ for $1<l<n$.
The minimal such $D$ is called the \emph{separable rank} of $\rho$, denoted $\textrm{sep-rank}( \rho) $. 
\end{definition}

With the construction from \cref{rem:trtompdo} it is clear that a state is separable if and only if it admits a separable decomposition. 

We finish by establishing a basic inequality of the separable rank---the proof is exactly as the one of \cref{prop:osrineq}.

\begin{proposition}
Let $0\leqslant \rho,\rho' \in \mc{H}^{[1]} \otimes \cdots \otimes \mc{H}^{[n]} $ be separable. Then so is $\rho+\rho',$ and
$$\textrm{sep-rank}(\rho+\rho')\leq\textrm{sep-rank}(\rho)+\textrm{sep-rank}(\rho').$$
\end{proposition}

%%============================================
\subsection{The local purification form}
\label{ssec:lp}

In this subsection we present and analyse another natural decomposition of $\rho$, namely the local purification, whose main feature is the fact that the local tensors are psd.  
The local purification form was introduced in Refs.\ \cite{Ve04d,Zw04}; 
results presented below can thus be found or be easily derived from Refs.\ \cite{Ve04d,Zw04}. 
The quantum square root rank (\cref{def:qsqrtrank}) is, to the best of our knowledge, introduced here for the first time.

To introduce this form, recall that the local physical space $\mc{H}^{[l]}$ is identified with $\M_d$. We denote the column space by  $\mc{V}^{[l]}$ and the row space by $\mc{V}^{[l]*}$, so that $\M_d =\mc{V}^{[l]*}\otimes  \mc{V}^{[l]}$. 
We also introduce an \emph{auxiliary} space associated to site $l$ as 
$ \mc{V}^{[l^\textrm{a}]}  = \C^{r}$ with some  $1\leq r$.

\begin{definition}{\rm \bf Local purification form}\cite{Ve04d}  \label{def:localpurif} 
Let $0\leqslant \rho \in \mc{H}^{[1]}\otimes \cdots \otimes \mc{H}^{[n]}$. 
A \emph{local purification form} of $\rho$  is defined as $\rho =  L L^\dagger $,  where $L$ is in Matrix Product Operator form (\cref{rem:mpo}), 
\be
L = \sum_{\alpha_1,\ldots,\alpha_{n-1}=1}^{D}
C^{[1]}_{\alpha_1}\otimes
C^{[2]}_{\alpha_1,\alpha_2}\otimes\cdots \otimes
C^{[n]}_{\alpha_{n-1}} ,
\label{eq:Lpurif}
\ee
where $C^{[l]}_{\alpha} \in \mc{V}^{[l]} \otimes  \mc{V}^{[l^\textrm{a}]*}$ for $l=1,n$  and 
$C^{[l]}_{\alpha,\beta} \in \mc{V}^{[l]} \otimes  \mc{V}^{[l^\textrm{a}]*}$ for $1<l<n$. 
The minimum such  $D$ is called the \emph{purification rank}, denoted $\purirank(\rho)$. Explicitly, 
$$
\purirank(\rho) = \min\{\osr(L)|LL^\dagger =\rho\}. 
$$
\end{definition}

Note that if the auxiliary space has the same dimension as the physical space, then simply $C_{\alpha,\beta}^{[l]} \in \mathcal M_d$. On the other hand, if $\rho$ is a pure state and thereby has $\rank(\rho)=1$, then $L$ is a column vector and $ \mc{V}^{[l^\textrm{a}]*}$ has dimension 1 for all $l^\textrm{a}$.

A local purification form always exists.  To see this,  
denote the spectral decomposition of   $\rho$ by $\rho = \sum_{j=1}^r \lambda_j |\psi_j\ra \la\psi_j| $, where $r=\rank(\rho)$ 
and define 
$$
L = \sum_{j=1}^r \sqrt{\lambda_j} |\psi_j\ra \la v_j|,
$$ 
where $\{|v_j\ra\}$ is some orthonormal basis. Then it is clear that $LL^\dagger =\rho$. 
In fact this fully characterises the set of $L$ such that $LL^\dagger=\rho$, that is, the only freedom is in the choice of the orthonormal basis $\{\la v_j|\}$. 
Therefore any such $L$ can be written as $L=L_0W^\dagger$, where $L_0$ is defined with, say, the computational basis, 
$$
L_0 = \sum_{j=1}^r \sqrt{\lambda_j} |\psi_j\ra \la j|, 
$$
and $W$ is an isometry,  $W : \mathbb{C}^{r} \to  \mathbb{C}^{r'}$ with $r'\geq r$, with $W^\dagger W =I$. 
This allows us to rewrite the purification rank as 
$$
\purirank(\rho) = \min_W\{\osr(L_0W^\dagger) | W^\dagger W =I \} .
$$ 
In words, the optimal local purification form will be given by the orthonormal basis $\{|v_j\ra\}$ that minimises $\osr(L)$.

\begin{remark}[A purification in the physics literature]
In the physics literature, a vectorised version of $L$ is called a purification. 
Explicitly, if $L$ is the matrix $L = \sum_{i,j}\ell_{ij}|i\ra\la j|$,  its vectorised version is denoted $|L\ra = \sum_{i,j} \ell_{ij}|i\ra |j\ra$. Denoting the second subsystem on which $|L\ra$ is defined as $\textrm{aux}$ (the auxiliary subsystem), we find that 
 $\rho = LL^\dagger = \tr_{\textrm{aux}}|L\ra\la L|$, where $\tr_{\textrm{aux}}$ is the partial trace over the auxiliary subsystem. 
\demo\end{remark}

Note  that the local purification form is not asking that $\rho$ has form  \eqref{eq:mpdo} with $A^{[l]}_{\alpha,\alpha'}\geqslant 0 $ for every $\alpha,\alpha'$. The latter is precisely the separable form, which exists only if $\rho$ is separable (see \cref{ssec:sepdecomp}). Instead, the local purification form always exists, and one has to be slightly more careful to see how the local matrices are psd. Namely, in the local purification form, 
\be
\quad\rho = \sum_{\alpha_1,\ldots,\alpha_{n-1},\beta_1,\ldots,\beta_{n-1}=1}^{D} 
B^{[1]}_{\alpha_1,\beta_1}
\otimes
B^{[2]}_{\alpha_1,\alpha_2,\beta_1,\beta_2}\otimes
\cdots \otimes 
B^{[n]}_{\alpha_{n-1},\beta_{n-1}}, 
\label{eq:lp}
\ee
where the $B$'s are psd matrices with respect to the following grouping of the indices:
\be  
\nonumber B^{[l]} &=&
\sum_{i,\alpha,j,\beta}
(B^{[l]})^{i,j}_{\alpha,\beta}
 |i,\alpha\ra 
 \la j,\beta| \geqslant 0 , 
\quad l=1,n\\
\nonumber B^{[l]} &= &
\sum_{i,\alpha,\alpha',j,\beta,\beta'}
(B^{[l]})^{i,j}_{\alpha,\alpha',\beta,\beta'}
 |i,\alpha,\alpha'\ra 
 \la j,\beta,\beta'| \geqslant 0,  \quad 1<l<n.
 \label{eq:Bl} 
\ee 
Explicitly, $B^{[l]}$ is constructed as $C^{[l]} C^{[l] \dagger}$, where $C^{[l]}$ are the local matrices of $L$ (Eq.\ \eqref{eq:Lpurif}), namely 
\begin{align*}
&B^{[l]}_{\alpha,\beta} = 
\sum_{i,j=1}^{d}
\sum_{k=1}^{r}(C^{[l]})^{i,k}_{\alpha}(\bar C^{[l]})^{k,j}_{\beta} |i\ra\la j|
, \quad l=1, n \\
&
B^{[l]}_{\alpha,\beta,\alpha',\beta'} = 
\sum_{i,j=1}^{d}
\sum_{k=1}^{r} 
(C^{[l]})^{i,k}_{\alpha,\alpha'} (\bar C^{[l]})^{k,j}_{\beta,\beta'}|i\ra\la j| , \quad  1<l<n .
\end{align*} 
Thus, this form contains a local certificate of positivity, in the sense that if the tensors $C^{[l]}$ are contracted as specified above, then $\rho\geqslant 0$ by construction. 
The problem is that the purification rank may need to be much larger than the operator Schmidt rank, as we will see in  \cref{thm:separation}. 

\begin{remark}{\rm \bf (Operational interpretation of the purification rank)}\cite{Ja13}\label{rem:operationalpuri}
The \emph{quantum correlation complexity} of a quantum state $\rho$, $\textrm{Q}(\rho)$, is defined as the minimum size of a seed that Alice and Bob need to share in order to produce $\rho$ via local operations \cite{Ja13}. The \emph{size} of a state is defined as half the number of qubits of the state \cite{Ja13}. In our terminology,
\be
\textrm{Q}(\rho) = \min\{t \mid &&\exists \: 0\leqslant \tau\in \mc{M}_{2^t} \otimes  \mc{M}_{2^t}\textrm{ and completely positive}\nn \\
&& \textrm{trace preserving maps}\:   \mc{E}_A, \mc{E}_B : (\mc{E}_A\otimes \mc{E}_B)(\tau) =\rho \}
\nn
\ee
Note that in the definition of $\textrm{Q}(\rho)$, Alice and Bob are allowed to do local operations but are not allowed to communicate (even classically). 
Thus, $\textrm{Q}(\rho)$ is nonincreasing under local operators, but not under classical communications, and therefore is only an upper bound  to the amount of entanglement. 

In addition, the \emph{quantum communication complexity} of a quantum state $\rho$, $\textrm{QComm}(\rho)$, is defined as the minimum number of qubits exchanged between Alice and Bob, initially sharing a product state, to produce $\rho$ at the end of the protocol.
Ref.\ \cite{Ja13} shows that  (without using the term purification rank) 
$$
\textrm{Q}(\rho) = \textrm{QComm}(\rho) = \lceil \log_2 \purirank(\rho)\rceil  .
$$ 
This thus gives an operational interpretation of  $\purirank$.

In \cref{rem:nonneg} we will comment on the operational interpretation of two related ranks, namely the nonnegative rank and the psd rank. 
Note also that a  multipartite version of the quantum correlation and quantum communication complexity is proposed in Ref.\ \cite{Ja14}, but with a different structure than the one considered here.
\demo\end{remark}

A basic inequality for the purification rank is stated in the following proposition. 
\begin{proposition}\label{pro:submulti}
For $0\leqslant \rho , \rho' \in \mc{H}^{[1]}\otimes \cdots \otimes \mc{H}^{[n]}$ we have
$$\purirank(\rho+\rho')\leq \purirank(\rho)+ \purirank(\rho').$$
\end{proposition}

%\begin{proof}  
\emph{Proof.}
Let $L$ and $L'$ be optimal local purifications of $\rho$ and $\rho'$, respectively.  We can append zero columns to all local matrices in $L$ and prepend zero columns to all local matrices in $L'$ without changing $\osr(L), \osr(L')$ and the fact that $LL^\dagger=\rho, L'L'^{\dagger}=\rho'$.  We can thus assume $LL'^\dagger=0=L'L^\dagger$. Then $L+L'$ provides a purification of $\rho+\rho'$ and we obtain \begin{align*}\purirank(\rho+\rho')&\leq \osr(L+L')\\ &\leq \osr(L)+\osr(L')\\ &=\purirank(\rho)+\purirank(\rho').%\qedhere
\end{align*}
\demo
%\end{proof}

%%--------------- 

To end this section, we introduce the quantum square root rank of a psd matrix. As we will see in \cref{thm:corresp}, this is the psd analogue of the square root rank of a nonnegative matrix.

\begin{definition}[The quantum square root rank]\label{def:qsqrtrank}
Let  $0\leqslant \rho\in \mc{H}^{[1]}\otimes \cdots \otimes \mc{H}^{[n]}$. 
The \emph{quantum square root rank} of $\rho$, denoted $\qsqrtrank(\rho) $, is defined as 
$$
\qsqrtrank(\rho) = \min_\tau\{\osr(\tau) | \tau^2 = \rho, \tau\  {\rm Hermitian} \} , 
$$ 
\end{definition}
Note that  the minimum is over  all Hermitian square roots of $\rho$. 
If we denote the spectral decomposition of $\rho$ by $\rho = U D U^\dagger$ with $D=\diag(\lambda_1, \lambda_2,\ldots)$, 
then its square roots are of the form 
$$
\tau = U D^{1/2} U^\dagger, \quad D^{1/2} =\diag(\pm\lambda_1, \pm\lambda_2,\ldots).
$$ 
Thus, $\qsqrtrank(\rho)$ is obtained by choosing the signs of the eigenvalues which minimise the operator Schmidt rank of $\tau$. 
It is obvious that the quantum square root rank upper bounds the purification rank,  see also \cref{pro:purisqrt}.

%%=========== translationally invariant case ====================

\section{Decompositions of t.i.\ psd matrices}
\label{sec:posti}

In this section we introduce and analyse decompositions of translationally invariant (t.i.) psd matrices in one spatial dimension.
We will first present general notions of a t.i.\ state (\cref{ssec:ti}), 
and then present the relevant decompositions in the t.i.\ case: 
the t.i.\ MPDO form (\cref{ssec:tiMPDO}),
the the t.i.\ separable decomposition (\cref{ssec:tisepdecomp}),
and the t.i.\ local purification form (\cref{ssec:tilp}).

\subsection{General notions}
\label{ssec:ti}
 For t.i.\ psd matrices $\rho$
we will denote the local Hilbert space associated to any individual subsystem by $\mc{H}_{\rm l}$, which is given by $\M_d$,
and the total Hilbert space by $\mc{H} =  (\mc{H}_{\rm l})^{\otimes n}$. Sometimes we will emphasize the system size by writing $\rho_n$ instead of $\rho$.  We start by defining translational invariance.

\begin{definition}[Translational invariance]\label{def:ti}
Let $0\leqslant \rho \in \mc{H} =  (\mc{H_{\rm l}})^{\otimes n}$. 
We say that $\rho$ is \emph{translationally invariant} (t.i.) if $T\rho T^\dagger =\rho$, where $T$ is the translation operator,
$$
T = \sum_{i_1,\ldots, i_n=1}^d |i_2,i_3,\ldots, i_1\ra \la i_1, i_2,\ldots, i_n|. 
$$
\end{definition}
Note that the action of $T$ just transforms an elementary tensor $A^{[1]}\otimes\cdots\otimes A^{[n]}$ to $A^{[2]}\otimes\cdots\otimes A^{[n]}\otimes A^{[1]}.$
So $\rho$ is t.i.\ if and only if it is invariant under cyclic permutations of the indices. Except for the case $n=2$, this is a weaker condition than being invariant under arbitrary permutations $\pi$ (i.e.\ $\rho = \pi\rho\pi^\dagger$), in which case $\rho$ would be called symmetric. 
Note also that if $\rho$ is t.i., then $\rho$ needs to have periodic boundary conditions, whereas in \cref{sec:pos} we considered  open boundary conditions. 
 
%%=============================
\subsection{The  t.i.\ MPDO form}
\label{ssec:tiMPDO}
We now define the t.i.\ analogue of the MPDO form. 
This form was considered already in \cite{Pe07} (although it was not given the name t.i.\ MPDO form, nor its associated rank was called t.i.\ osr). We will prove some slightly more general result in \cref{pro:W}, as well as the easily derivable properties of \cref{pro:tiosr}.

\begin{definition}[T.i.\ MPDO]
\label{def:tiMPDO}
Let $0\leqslant \rho \in (\mc{H}_{\rm l})^{\otimes n}$. 
A \emph{translationally invariant (t.i.) MPDO} form of $\rho$ is given by
\be
\rho = \sum_{\alpha_1,\ldots,\alpha_{n}=1}^D 
A_{\alpha_1,\alpha_2} \otimes 
A_{\alpha_2,\alpha_3} \otimes 
\cdots \otimes
A_{\alpha_{n},\alpha_1} ,
\label{eq:timpdo}
\ee
where $A_{\alpha,\alpha'}\in\mc{H}_\mathrm{l}$. 
The minimal such $D$ is called the t.i.\ operator Schmidt rank of $\rho$, denoted $\tiosr(\rho)$.
\end{definition}

Note that in the case $n=2$, the t.i.\ MPDO form, as we have defined it here, is  $$\rho=\sum_{\alpha,\beta=1}^D A_{\alpha,\beta}\otimes A_{\beta,\alpha}.$$ We will restrict to this kind of decomposition throughout this section, and will go back to this observation in  \cref{ssec:symm}.

Clearly, a t.i.\ representation such as \eqref{eq:timpdo} implies that $\rho$ is t.i. 
The converse is also true, but generally not at a fixed cost, i.e.\ $D$ needs to grow with $n$:  

\begin{remark}{\rm \bf (Imposing translational invariance)}\cite{Pe07} \label{rem:ti} If $\rho$ is t.i.\ and has a non-t.i.\ representation, say an MPDO form given by \eqref{eq:mpdo}, this can be made t.i.\ at the expense of increasing $D$ to  $Dn$, in the general case. 
To see this, first note that we can transform  \eqref{eq:mpdo} to $$\rho=\sum_{\alpha_1,\ldots, \alpha_{n}=1}^D A_{\alpha_1,\alpha_2}^{[1]}\otimes  A_{\alpha_2,\alpha_3}^{[2]}\otimes\cdots\otimes A_{\alpha_{n-1},\alpha_n}^{[n]}$$ by just padding the row vector $A^{[1]}$ and the column vector $A^{[n]}$  with zeros (we express the virtual indices here). Then define $$A= n^{-1/n}\left(\begin{array}{cccc}0 & A^{[1]} & 0 & 0 \\0 & 0 & \ddots & 0 \\0 & 0 & 0 & A^{[n-1]} \\A^{[n]} & 0 & 0 & 0\end{array}\right).$$ Since $\rho$ is t.i.\ one immediately verifies that $$\rho=\sum_{\alpha_1,\ldots,\alpha_n=1}^{Dn} A_{\alpha_1,\alpha_2}\otimes \cdots\otimes A_{\alpha_{n-1},\alpha_n}.$$ 
\demo\end{remark}

We now consider a state which is t.i.\ but in a non-trivial way.

\begin{example}[The $W$ state] \label{ex:Wstate}
Consider a  pure state $\rho = |W\ra\la W|$, 
so that by \cref{pro:pure} it suffices to study the operator Schmidt rank of $|W\ra$.   
Here  $|W\ra$ is the so-called $W$ state on $n$ sites, 
\be
|W\ra = n^{-1/2} 
\sum_{j=1}^n\sigma_x^{(j)}
|0\ra^{\otimes n} . \label{eq:W}
\ee
Here $\sigma_x^{(j)}$ denotes the operator $\sigma_{x}=|0\ra\la 1| + |1\ra\la 0|$ acting on site $j$, and $|0\ra^{\otimes n}$ denotes the $n$-fold tensor product of $|0\ra$. For example, for $n=3$, 
$$
|W\ra = \frac{1}{\sqrt{3}}( |0,0,1\ra+|0,1,0\ra +|1,0,0\ra).
$$
 This state has a  non-t.i.\ representation with $D=2$ given by 
\be \label{eq:ABW}
A^0 = \begin{pmatrix} 1&0\\ 0&1\end{pmatrix}, \quad A^1 = \begin{pmatrix} 0&1\\ 0&0\end{pmatrix},\quad 
B =  \begin{pmatrix} 0&0\\ 1&0\end{pmatrix}, 
\ee
namely 
\be
|W\ra = n^{-1/2} \sum_{i_1,\ldots, i_n=0}^1 \tr(B A^{i_1}A^{i_2}\cdots A^{i_n})|i_1,\ldots,i_n\ra .
\label{eq:Wnti}
\ee
We can obtain a t.i.\ representation of size $2n$ by  using the idea of \cref{rem:ti}. 
Explicitly, for a given $n$ we define
\be
C^i = n^{-1/n}
\begin{pmatrix}
0 & BA^{i} &0 & \ldots &0\\
0 & 0 & A^{i} &\ldots & 0\\
\vdots & \vdots &\ddots &  \ddots &\vdots\\
A^{i} & 0 & \ldots & 0& 0
\end{pmatrix}
\label{eq:C}
\ee
and we have 
\be
|W\ra =\sum_{i_1,\ldots, i_n=0}^1 \tr(C^{i_1}C^{i_2}\cdots C^{i_n})|i_1,\ldots,i_n\ra.
\label{eq:Wti}
\ee
Note that the size of $C^i$ is $2n$.
\demo\end{example}

%%---------------------------------------
We now show that  any t.i.\ representation of the $W$ state requires a bond dimension which grows at least as $\sqrt{n}$, where $n$ is the system size.

\begin{proposition} \label{pro:W}
Let $|W\ra$ be the $W$ state on $n$ sites defined in \cref{ex:Wstate}. Then 
$$
\tiosr(|W\ra) \geq \sqrt{n}.
$$ 
\end{proposition}

A proof that $\tiosr(|W\ra) \geq \Omega(n^{1/3})$ is provided in \cite[Appendix]{Pe07} and the subsequent proof of Wielandt's Theorem \cite{Sa10} (see \cite{Mi18b} for the latest, stronger result).
Here we use the results of Ref.\ \cite{De17} to prove this bound which is, to the best of our knowledge, new, and independent of the quantum Wielandt's Theorem.

%\begin{proof}
\emph{Proof.}
First observe that $|W\ra$ is t.i.\ in a non-trivial way, as it contains a sum of $n$ states, each of which is invariant under $T^n$. The translation operator $T$ on $|W\ra$ thus generates a cyclic permutation of these terms. 
In the language of Ref.\ \cite{De17}, $|W\ra$ is $n$-periodic. 

Now consider a t.i.\ representation of $|W\ra$ such as the one of \eqref{eq:Wti}. 
From  Ref.\ \cite{De17} it follows that
the transfer matrix $E := \sum_{i=0}^1 C^i\otimes \bar C^i$ must have $n$ eigenvalues of modulus 1 of the form 
$$
\{e^{i2\pi r /n}\}_{r=0}^{n-1}.
$$ 
Note that $E$ is a matrix of size $D^2$. 
But $n\leq D^2$ and therefore $D\geq \sqrt{n}$. 
\demo
%\end{proof}

\begin{remark}[T.i.\ Matrix Product Operator form]\label{rem:timpo}
In analogy to \cref{rem:mpo}, given a t.i.\ operator 
$L\in \mc{M}_{d}\otimes \cdots \otimes  \mc{M}_{d}$ ($n$ times), 
we define its t.i.\ Matrix Product Operator form as a decomposition
\be
L = \sum_{\alpha_1,\ldots,\alpha_{n}=1}^{D}
C_{\alpha_1,\alpha_2}\otimes
C_{\alpha_1,\alpha_2}\otimes\cdots \otimes
C_{\alpha_{n},\alpha_1} ,
\label{eq:Lpurif}
\ee
where $C_{\alpha,\beta} \in \mc{M}_{d}$. 
The minimum such $D$ is also called the t.i.\ operator Schmidt rank of $L$, $\tiosr(L)$. 

The situation is entirely parallel to the non-t.i.\ case: the only difference between the t.i.\ MPO and t.i.\ MPDO form is that in the latter the operator is globally psd. 
\demo\end{remark}

The following result is again proven analogously to \cref{prop:osrineq}.
\begin{proposition} \label{pro:tiosr}
Let $\rho,\tau \in (\mc{H}_{\rm l})^{\otimes n}$ be t.i. Then so are $\rho+\tau, \rho\tau$ and
\begin{itemize}
\item[(i)] $\tiosr(\rho+\tau)\leq\tiosr(\rho)+\tiosr(\tau).$
\item[(ii)] $\tiosr(\rho\tau)\leq\tiosr(\rho)\tiosr(\tau).$
\end{itemize}
\end{proposition}

%%============================
\subsection{The t.i.\ separable decomposition}
\label{ssec:tisepdecomp}

Here we introduce and characterise the t.i.\ separable decomposition, which, to the best of our knowledge, is new. 
Before introducing the t.i.\ separable decomposition, we start by defining and characterising  t.i.\ separable psd matrices.

\begin{definition}[T.i.\ separable psd matrix]\label{def:tisep}
Let $0\leqslant \rho \in (\mc{H}_\mathrm{l})^{\otimes n}$. 
We say that $\rho$ is \emph{t.i.\ separable} if $\rho$ is t.i.\ (\cref{def:ti}) 
and $\rho$ is separable (\cref{def:sep}).
\end{definition}

It is easy to see that $\rho$ is t.i.\ separable if and only if it is of the form 
$$
\rho = \frac{1}{n} \sum_{m=1}^{n} T^m \sigma T^{\dagger m}
$$
where $\sigma$ is separable.
Note that if $\rho$ is a sum of t.i.\ product matrices
$$
\rho = \sum_{\alpha}  \sigma_\alpha \otimes \cdots \otimes \sigma_\alpha
$$
where $\sigma_\alpha\geqslant 0$,  
then  $\rho$ is t.i.\ separable, but the converse is not true, as the following example shows: 
\be
\rho = \frac{1}{2} (|0\ra \la 0| \otimes |1\ra \la 1| + |1\ra \la 1|\otimes |0\ra \la 0|). 
\label{eq:mixedW2}
\ee
See \cref{ex:modifiedW} for more properties of this state.

We now define the t.i.\ separable decomposition. 

\begin{definition}[T.i.\ separable decomposition]\label{def:tisepdecomp}
Let $0\leqslant \rho \in (\mc{H}_\mathrm{l})^{\otimes n}$. 
A \emph{t.i.\ separable decomposition} of $\rho$ is a form 
\be
\label{eq:tisepdecomp}
\rho = \sum_{\alpha_1,\ldots, \alpha_n=1}^D \chi_{\alpha_{1},\alpha_2} \otimes
\chi_{\alpha_{2},\alpha_3}  \otimes \cdots \otimes 
\chi_{\alpha_{n},\alpha_1}  , 
\ee
where $\chi_{\alpha,\beta} \geqslant 0 $ for all $\alpha,\beta$. 
The minimal such $D$ is called the \emph{t.i.\ separable rank} of $\rho$, denoted $\textrm{ti-sep-rank}( \rho) $. 
\end{definition}

This definition captures precisely the set of t.i.\ separable psd matrices: 

\begin{proposition} \label{lem:tiseptidecomp}
Let $0\leqslant \rho \in (\mc{H}_\textrm{l})^{\otimes n}$. 
Then $\rho$ is t.i.\ separable (\cref{def:tisep}) if and only if it admits a t.i.\ separable decomposition (\cref{def:tisepdecomp}). 
\end{proposition}

%\begin{proof}
\emph{Proof.}
Let $\rho$ be  t.i.\ separable. If we write down a separable decomposition of $\rho$ and then apply the construction from \cref{rem:ti} we obtain the desired t.i.\ separable decomposition.
The converse direction is clear.
\demo
%\end{proof}

The following  example is a modified version of the $W$ state (\cref{ex:Wstate}).

\begin{example}[Mixed state version of the $W$ state]\label{ex:modifiedW}
Consider the t.i.\ separable state
$$
\rho = \frac{1}{n} \sum_{i=1}^n \sigma_x^{(i)} (|0\ra\la 0|)^{\otimes n}\sigma_x^{(i)}. 
$$
Note that Eq.\  \eqref{eq:mixedW2} corresponds to this state for $n=2$, and thus this state is not a convex combination of t.i.\ product states. 
We want to see that  
\be
\tiseprank(\rho)\geq \frac{\sqrt{n}}{2} \: \seprank(\rho) =  \sqrt{n} .
\label{eq:tisepsep}
\ee 
We use essentially the same argument as in \cref{pro:W}. 

We first provide a separable decomposition of $\rho$ with $\seprank(\rho)=2$. 
Define the 4-tensor $\chi = \{\chi_{\alpha,\beta}^{i,j}\}$ using the definitions of \eqref{eq:ABW}: 
$\chi^{0,0} = A^0$, 
$\chi^{1,1} = A^1$, and the rest 0, 
as well as $B$ defined there. 
Then 
$$
(\tau^{[1]})^{i,j} = B \chi^{i,j}, \quad (\tau^{[l]})^{i,j} = \chi^{i,j}\quad 1<l\leq n
$$
provide a separable decomposition of $\rho$ with $\seprank(\rho)=2$.

Now we want to see the first inequality of  \eqref{eq:tisepsep}. So consider an optimal t.i.\ separable decomposition given by a tensor 
$\tau = \{\tau_{\alpha,\beta}^{i,j}\}$. 
By construction $\tau_{\alpha,\beta} \in \mc{M}_2$ is psd for all $\alpha,\beta$, and $D=\tiseprank(\rho)$.
First note that since $\rho$ is diagonal in the computational basis, we can assume that 
$\tau_{\alpha,\beta}^{i,j} = \delta(i,j) C^i_{\alpha,\beta} $ 
for some $C^i_{\alpha,\beta}$. 
But 
$$
\sum_{i_1,\ldots, i_n=0}^1 \tr(C^{i_1}\cdots C^{i_n})|i_1,\ldots, i_n\ra = |W\ra
$$
is the $W$ state,  and \cref{pro:W} shows that the bond dimension of  $C$ is $\geq \sqrt{n} $. 
\demo\end{example}

The next result is again proven analogously to  \cref{prop:osrineq}.

\begin{proposition}
Let $0\leqslant \rho,\rho' \in (\mc{H}_\mathrm{l})^{\otimes n}$ be t.i.\ separable. Then so is $\rho+\rho'$ and $$\textrm{ti-sep-rank}(\rho+\rho')\leq\textrm{ti-sep-rank}(\rho)+ \textrm{ti-sep-rank}(\rho').$$
\end{proposition}

%%===================================
\subsection{The t.i.\ local purification form}
\label{ssec:tilp}

Here we characterise the t.i.\ local purification form, which had been considered, e.g. in \cite{De15} (although it was not given this name). 
Before introducing the t.i.\ separable decomposition, we start by defining and characterising  t.i.\ separable psd matrices. 
In the t.i.\ MPDO form of \cref{def:tiMPDO}, the local tensors $A_{\alpha,\beta}$ need not be psd, as in the MPDO form. Enforcing positivity on the local matrices leads to the t.i.\ local purification form. 

To introduce it, we define 
$\mc{V}_\textrm{l} $ as the column space associated to the local, physical Hilbert space $\mc{H}_{\textrm{l}}$, and 
$\mc{V}^\textrm{a}_\textrm{l}$  as the column space associated to the local, auxiliary Hilbert space $\mc{H}^{\textrm{a}}_{\textrm{l}}$.

\begin{definition}[T.i.\ local purification]
\label{def:tilocalpurif} 
Let $0\leqslant \rho \in (\mc{H}_\mathrm{l})^{\otimes n}$. 
A \emph{t.i.\ local purification form} of $\rho$  is defined as $\rho = LL^\dagger $,  where  $L$ is in t.i.\ Matrix Product Operator form (\cref{rem:timpo}), 
\be
\label{eq:Ltilp}
L= 
\sum_{\alpha_1,\ldots,\alpha_{n}=1}^{D}
C_{\alpha_1,\alpha_2}\otimes
C_{\alpha_2,\alpha_3}\otimes
\cdots \otimes
C_{\alpha_{n},\alpha_1} ,
\ee
where $C_{\alpha,\beta} \in \mc{V}_\textrm{l} \otimes \mc{V}^\textrm{a}_\textrm{l}$. 
The minimum such $D$ is called the t.i.\ purification rank, denoted $\tipurirank(\rho)$. Explicitly, 
$$
\tipurirank(\rho) = \min\{\tiosr(L) | LL^\dagger = \rho\}. 
$$
\end{definition}

\begin{remark}[Existence of the t.i.\ local purification]\label{rem:exlocpuri}
Note that a t.i.\ local purification exists for every t.i.\ psd matrix. The unique psd square-root of $\rho$ is a polynomial expression in $\rho$, and thus also t.i. It therefore  admits a t.i.\  MPO form, as argued in \cref{rem:ti}. This provides a t.i.\ local purification of $\rho$ (which is generally not the optimal one).
\demo\end{remark}

To see how the local matrices of the t.i.\ local purification are psd, we proceed similarly as in  \cref{ssec:lp}. Namely, $\rho$ has the form 
\be
\qquad
\rho = \sum_{\alpha_1,\ldots,\alpha_{n},\beta_1,\ldots,\beta_{n}=1}^{D} 
B_{\alpha_1,\alpha_2,\beta_1,\beta_2}\otimes 
B_{\alpha_2,\alpha_3,\beta_2,\beta_3} \otimes 
\cdots \otimes 
B_{\alpha_{n},\alpha_1,\beta_n,\beta_1} ,
\label{eq:lpti}
\ee
where $B$ is psd because it is constructed as $CC^\dagger$, where $C$ is given by  Eq.\ \eqref{eq:Ltilp}; explicitly: 
$$
%\label{eq:Bti}
 B = \sum_{i,j=1}^d \sum_{\alpha,\alpha',\beta,\beta'=1}^D
 \sum_{k=1}^{r} 
 C^{k,i}_{\alpha,\alpha'} \bar C^{j,k}_{\beta,\beta'} |i,\alpha,\alpha'\ra\la j,\beta,\beta'| \geqslant 0. 
$$

The proof of the following inequality is similar to the one of \cref{pro:submulti}.

\begin{proposition}
Let $0\leqslant \rho,\rho' \in (\mc{H}_\mathrm{l})^{\otimes n}$ be t.i. Then so is  $\rho+\rho'$ and
$$\tipurirank(\rho+\rho')\leq \tipurirank(\rho)+ \tipurirank(\rho').$$
\end{proposition}

%%===========================
\section{Correspondence with factorisations of nonnegative matrices}
\label{sec:matrixcones}

In this section  we  present a correspondence of  decompositions of bipartite psd matrices which are diagonal in the computational basis with factorisations of nonnegative matrices.  
The results of this section are new, except when we review definitions (as in \cref{ssec:decomp}). 
First we will comment on the difference between t.i.\ decompositions and symmetric decompositions in the bipartite case (\cref{ssec:symm}), since in this section we will be interested in symmetric bipartite decompositions. 
Then we will define the factorisations of nonnegative matrices relevant for our problem (\cref{ssec:decomp}), 
and finally we will present the correspondence  (\cref{ssec:correspondence}).

%%=======================
%%=======================
\subsection{Comparison with symmetric decompositions}
\label{ssec:symm}
In \cref{sec:pos} we analysed the $\osr$, the $\seprank$ and the $\purirank$, and  
 in  \cref{sec:posti} their t.i.\ analogues, namely the $\tiosr$, the $\tiseprank$ and the $\tipurirank$. 
Another natural decomposition is the fully symmetric version of each of the three ranks, 
where for example the symmetric tensor rank \cite{Co08c} is defined to be the minimal $r$ such that 
$$
\rho = \sum_{\alpha=1}^r A_\alpha \otimes A_\alpha \otimes \ldots \otimes A_\alpha . 
$$
Similarly, one could define the symmetric $\seprank$ and the symmetric  $\purirank$. 
Now, although full symmetry and translational invariance coincide in the bipartite case,  
the symmetric and the t.i. decompositions do not coincide. 
Namely, the symmetric decomposition results in 
  \be 
\label{eq:specialsymm}
\rho=\sum_{\alpha} A_{\alpha}\otimes A_{\alpha}, 
\ee
whereas the t.i.\ decomposition results in 
\be
\label{eq:specialti}
\rho=\sum_{\alpha,\beta} A_{\alpha,\beta}\otimes A_{\beta,\alpha}.
\ee
In this section we will be interested in decompositions of type \eqref{eq:specialsymm}. For this reason we now include some general existence result about symmetric decompositions. 

\begin{proposition}\label{prop:exspecial}
Every t.i.\ $0\leqslant \rho \in \mc{H}_{\rm l}\otimes \mc{H}_{\rm l}$  admits a decomposition of the form \eqref{eq:specialsymm}, 
and a t.i.\ local purification of the form \eqref{eq:specialsymm}. 
\end{proposition}

However, this is not the case for the separable decomposition, as we will see in \cref{cor:cpsdt}.

%\begin{proof}
\emph{Proof.}
In the bipartite case, translational invariance is the same as full symmetry, so the first statement is just the well-known decomposition of symmetric matrices \cite{Co08c}. 
In fact, the t.i.\ operator Schmidt rank equals the operator Schmidt rank. A non-optimal decomposition can also explicitly be obtained   by first choosing a decomposition on the double edge, 
$$
\rho=\sum_{\alpha,\beta}A_{\alpha,\beta}\otimes A_{\beta,\alpha}
$$ 
whose existence we have seen above, and using that 
\begin{align*}
\sum_{\alpha,\beta}(A_{\alpha,\beta}+A_{\beta,\alpha})\otimes (A_{\alpha,\beta}+A_{\beta,\alpha})+i(A_{\alpha,\beta}-A_{\beta,\alpha})\otimes i(A_{\alpha,\beta}-A_{\beta,\alpha})&=4\rho.
\end{align*} 
The existence of the t.i.\ local purification then follows from the argument of \cref{rem:exlocpuri}.
\demo
%\end{proof}

%%==================================
\subsection{Factorisations of nonnegative matrices}
\label{ssec:decomp}

We now consider a rectangular nonnegative matrix $M\in \mathbb{R}_{+}^{p\times q}$, 
 where $\mathbb{R}_{+}$ denotes the set of non-negative reals. 
 We will consider six factorisations of $M$, 
 and each will be associated to a roman number which we will use in  \cref{thm:corresp}.
 For the symmetric factorisations [(iv), (v), and (vi)], $M$ will need to be square and symmetric. 
 For every factorisation there will be a  minimal dimension of the matrices involved, which defines the rank associated to that factorisation. 
The factorisations are the following: 
\begin{itemize}
\item[(i)] A \emph{minimal factorisation} is an expression $M=AB$ where $A$ has $\rank(M)$ columns. 
If $M$ is real, $A$ and $B$ can be chosen real without loss of generality. 
This factorisation can be obtained, for instance, by doing the singular value decomposition of $M=U\Sigma V$ and absorbing the singular values somewhere. 

%--------------------------------------
\item[(ii)]
The  \emph{nonnegative factorisation} \cite{Ya91}  is an expression $M=AB$ where $A$ and $B$ are nonnegative, i.e.\ $A\in \mathbb{R}_+^{p\times r}$ and $B\in \mathbb{R}_+^{r\times q}$. The minimal such $r$ is called the nonnegative rank, denoted $\textrm{rank}_+(M)$.

%--------------------------------------
\item[(iii)]
The \emph{positive semidefinite (psd) factorisation} \cite{Fi11,Fa14} is an expression $M_{i,j} =\tr(E_iF_j^t)$, where $E_i$ and $F_j$ are psd matrices of size $r\times r$ with entries in the rationals, the real or the complex numbers.  
The minimal such $r$ is called the \emph{psd rank}, denoted 
$\psdrank^{\mathbb{Q}}(M)$, 
$\psdrank^{\mathbb{R}}(M)$, and 
$\psdrank^{\mathbb{C}}(M)$, respectively.

%%----------------------
\item[(iv)] 
A \emph{symmetric factorisation} is an expression $M= AA^t$,  where $A\in  \mathbb{C}^{p\times r}$. The minimal such $r$ is called the \emph{symmetric rank} of $M$, denoted $\symmrank(M)$. 
Such a decomposition can be found by diagonalisation of symmetric bilinear forms over $\C$ \cite{Co82}.
In fact, there always exists some invertible (complex) matrix $P$ such that $PMP^t$ is diagonal, with only ones and zeros on the diagonal. This can also be understood as doing elementary row {\it and the same} column operations to $M$ over $\C$, to bring it to diagonal form. If the resulting diagonal matrix has $r$ ones on the diagonal,  then  $A:=P^{-1}\left(\begin{array}{c}I_r \\0\end{array}\right)$ provides a symmetric decomposition. From this construction it also follows that  $\symmrank(M)=\rank(M)$ for symmetric matrices.

%--------------------------------------
\item[(v)]
The \emph{completely positive (cp) factorisation} \cite{Be15b}  is an expression $M=AA^t$, where $A$ is a nonnegative matrix $A\in \mathbb{R}_+^{p\times r}$. The minimal such $r$ is called the  \emph{cp rank} of $M$, denoted $\textrm{cp-rank}(M)$.

%--------------------------------------
\item[(vi)] 
The \emph{completely positive semidefinite transposed (cpsdt) factorisation}    is an expression $M_{i,j} =\tr(E_iE_j^t)$, where $E_i$ are psd matrices of size $r\times r$ with entries in the rationals, the reals or the complex numbers. 
The minimal such $r$ is called the \emph{cpsdt rank},  denoted 
$\cpsdrank^{\mathbb{Q}}(M)$,
$ \cpsdrank^{\mathbb{R}}(M)$, and 
$\cpsdrank^{\mathbb{C}}(M)$, respectively.
\end{itemize}

\begin{remark}
Note the transposition on the second term in the definition of a cpsdt factorisation. 
If the psd matrices $E_i$ have entries in  $\mathbb Q$ or $\mathbb R$ it can  be omitted, since such psd matrices are symmetric. 
So the notion of a cpsdt factorisation coincides with the notion of a \emph{completely positive semidefinite factorisation} (cpsd) \cite{La15c} in that case, which has been studied a lot recently. 
However, the transpose makes a difference over the complex numbers. 
Indeed, a  cpsd factorisation does not exist for every symmetric nonnegative matrix, for example because it requires that the matrix is positive semidefinite (and even this is not sufficient in general). 
However, a cpsdt factorisation does always exist, as we will see in \cref{cor:cpsdt}.

Also note that in the definition of the psd factorisation, one could omit the transposition at the $F_i$ without changing the notion, since $E_i$ and $F_i$ are independent anyway. 
\demo\end{remark}

\begin{figure}[htb]
\[ \begin{tikzcd}
\textrm{(i) minimal factorisation}  \arrow{r}{\textrm{symmetric}} 
\arrow[swap]{d}{\textrm{nonnegative}} 
& \textrm{(iv) symmetric factorisation} \arrow{d}{\textrm{nonnegative}} \\
\textrm{(ii) nonnegative factorisation} 
\arrow{r}{\textrm{symmetric}} 
\arrow[swap]{d}{\textrm{non-commutative}} 
& \textrm{(v) cp factorisation} \arrow{d}{\textrm{non-commutative}} \\
\textrm{(iii) psd factorisation}\arrow{r}{\textrm{symmetric}}& 
\textrm{(vi) cpsdt factorisation}
\end{tikzcd}
\]
\caption{The relations among the factorisations of nonnegative matrices.}
\label{fig:decomp}
\end{figure}
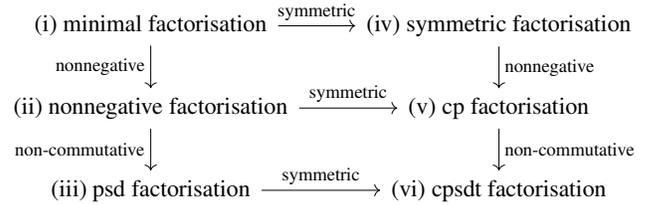

Note that the symmetric factorisation, 
the cp factorisation, and the cpsdt factorisation are the symmetric versions of 
the minimal factorisation, the nonnegative factorisation, and the psd factorisation, respectively. 
At the same time, 
the nonnegative factorisation and the cp factorisations can be obtained from the 
minimal factorisation and the symmetric factorisation, respectively, 
by imposing that the matrices are nonnegative.
Moreover, 
the psd and the cpsdt factorisations are the non-commutative generalisations of the 
nonnegative factorisation and the cp factorisation, respectively. These relations are summarised in Figure \ref{fig:decomp}.

%%--------------------

Clearly the symmetric decompositions [(iv), (v), and (vi)] can only exist for symmetric matrices (i.e.\ for  $M=M^t$). 
However, while every symmetric nonnegative matrix has a symmetric and a cpsdt factorisation, 
not every symmetric nonnegative matrix has a cp decomposition. One obvious reason is that if $M$ admits a cp factorisation, then $M$ needs to be psd as well---see \cite{La15c} or  \cite{Fa14}. 
This is not the case in the symmetric factorisation, since the entries of $A$ can be complex.

%%--------------------

Finally we review the definition of the square root rank of a nonnegative matrix (see, e.g., \cite{Fa14}). We will use the \emph{Hadamard square root decomposition}, which is an expression $M = N \circ N$, where  $\circ$ denotes the Hadamard product (i.e.\ entrywise multiplication, $n_{ij}^2 = m_{ij}$). Note that the square root rank is denoted $\rank_{\sqrt{}}$ in  \cite{Fa14}.

\begin{definition}[The square root rank]\label{def:sqrtrank}
Let $M$ be a nonnegative matrix. The \emph{square root rank} of $M$, denoted $\sqrtrank(M)$, is defined as 
$$
\sqrtrank(M) = \min \{\rank (N) | N\circ N = M\}.
$$
\end{definition}

The minimisation is thus over all entrywise square roots of $M$, i.e.\ for each entry we can choose the positive or the negative square root.

%%=====================================
We end this section by reviewing the operational interpretation of the nonnegative and the psd rank, as given in Ref.\ \cite{Ja13}.

\begin{remark}{\rm \bf (Operational interpretation of the nonnegative  and the psd rank)}\label{rem:nonneg}
Given a bipartite probability distribution $p=\{p(x,y)\}_{x,y}$, define its \emph{size} as half of the total number of bits. 
The \emph{randomized correlation complexity} $\textrm{R}(p)$ is defined as the minimum size of a joint probability distribution that Alice and Bob need to share to produce $p$ by means of only local operations \cite{Ja13}.
Ref.\ \cite{Ja13} also considers a  communication scenario in which Alice and Bob do not share anything from the start.
The \emph{randomized communication complexity} $\textrm{RComm}(p)$ is the minimum number of bits that they need exchange to produce $p$ \cite{Ja13}. 
 Then 
 $$\textrm{R}(p) = \textrm{RComm}(p) = \lceil \log_2 \rank_+ (p)\rceil .$$ 

Concerning the psd rank, given a bipartite quantum state $\rho$, define its \emph{size} as half of the total number of qubits. 
The \emph{quantum correlation complexity} $\textrm{Q}(p)$ is defined as the minimum size of a quantum state that Alice and Bob have to share to produce $p$ (Alice and Bob can then apply local operations) \cite{Ja13}.
Similarly, the \emph{quantum communication complexity} $\textrm{QComm}(p)$ is the minimum number of qubits that they have to send to each other to produce $p$ \cite{Ja13}. 
Then 
$$\textrm{Q}(p) = \textrm{QComm}(p) = \lceil \log_2 \psdrank(p)\rceil . $$

Another operational interpretation is provided by \cite{Fi11}. 
Given a joint probability distribution $p$, $\log(\rank_+ (p))$ and $\log(\psdrank (p))$ are used to characterise the amount of classical or, respectively, quantum communication needed to compute $p$ in expectation. 

A further interpretation is given in Ref.\ \cite{We08}, where the nonnegative matrix is a matrix of measurement outcomes on a quantum state, and the psd rank is used to determine the minimal dimension of the Hilbert space where the quantum state lives. 
\demo\end{remark}

%%==========================================
\subsection{Correspondence of decompositions}
\label{ssec:correspondence}

Here we present a correspondence between 
the factorisations of nonnegative matrices of \cref{ssec:decomp} and decompositions of bipartite psd matrices which are diagonal in the computational basis, namely 

\be
\sigma = \sum_{i=1}^{d_1} \sum_{j=1}^{d_2} m_{ij} |i,j\ra\la i,j|. 
\label{eq:cb}
\ee
We reserve the letter $\sigma$ for such psd matrices; a general psd matrix is called $\rho$. 
Clearly, $\sigma$ is psd if and only if $M = (m_{ij})_{i,j} \in \R_+^{d_1\times d_2}$ is a nonnegative matrix. 

We will relate the symmetric decompositions of $\sigma$ given in \cref{ssec:symm} with the symmetric factorisations  of $M$. 
In this case, $d_1=d_2=:d$ and we have that $\sigma$ is t.i.\ which  if and only if $M$ is symmetric, i.e.\ $M=M^t$.

\begin{theorem}\label{thm:corresp}
Consider the psd matrix $\sigma$ of \eqref{eq:cb}, and let 
$$M= \sum_{i=1}^{d_1} \sum_{j=1}^{d_2} m_{ij} |i\ra\la j|$$ denote the nonnegative matrix containing  the diagonal elements of $\sigma$. Then the following correspondence of decompositions, and consequently  of ranks, holds: 

\begin{table}[htb]\centering
\begin{tabular}{cr|l}
&Decomposition of $\sigma$ & Decomposition of $M$ \\ \hline
(i)&operator Schmidt decomposition & minimal factorisation \\
(ii)&separable decomposition&  nonnegative factorisation  \\
(iii)& local purification & complex psd factorisation \\
(iv)&t.i.\ operator Schmidt decomposition& symmetric factorisation  \\
(v)&t.i.\ separable decomposition  &  cp factorisation\\
(vi)&t.i.\ local purification & complex cpsdt factorisation \\
\end{tabular}
\label{tab:corresp}
\end{table}

\begin{table}[htb]\centering
\begin{tabular}{crcl}
&Corresponding rank of  $\sigma$  && Corresponding rank of   $M$ \\ \hline
(i)&$\osr( \sigma) $ &$=$& $\rank(M)$\\
(ii)&$\seprank( \sigma) $ &$=$& $\rank_+(M)$\\
(iii)&$\purirank (\sigma)$ &$=$& $ \psdrank^{\mathbb{C}}  (M)$\\
(iv)&$\tiosr( \sigma) $ &$=$& $\symmrank(M)$\\
(v)&$ \tiseprank( \sigma)$ &$=$& $\cprank(M)$\\
(vi)&$\tipurirank( \sigma)$ &$=$& $\cpsdrank^{\mathbb{C}} (M)$\\
\end{tabular}
\label{tab:ranks}
\end{table}
 
In addition, the following correspondence of ranks holds: 

 \begin{table}[htb]\centering
\begin{tabular}{crcl}
&Corresponding rank of  $\sigma$  && Corresponding rank of   $M$ \\ \hline
(vii) &$\qsqrtrank( \sigma) $ &$=$& $\sqrtrank(M)$ 
\end{tabular}
\label{tab:qsqrtrank}
\end{table}
\end{theorem}

Note that the $\tiosr$, $ \tiseprank$ and $\tipurirank$ refer to decompositions of type \eqref{eq:specialsymm}, as remarked in \cref{ssec:symm}.
The proof of \cref{thm:corresp} is given in \cref{app:proofmain}. We remark that correspondence (ii) was already observed in \cite{De13c} and \cite{Ja14}.

\begin{corollary}\label{cor:cpsdt}
(i) There are bipartite separable t.i.\ psd matrices that do not admit a symmetric   separable decomposition of the type \eqref{eq:specialsymm}.

(ii) There are bipartite t.i. psd matrices, diagonal in the computational basis, that do not admit a real symmetric local purification of type \eqref{eq:specialsymm}. 

(iii) Every symmetric nonnegative matrix admits a (complex) cpsdt-factorization.
\end{corollary}

%\begin{proof}
\emph{Proof.}
(i) It is well-known that not every symmetric nonnegative matrix has a cp factorization. For example, being psd is an additional necessary condition, but even this is not sufficient for $d\geq 5$.

(ii) As the proof of \cref{thm:corresp} shows,  a real symmetric local purification of type \eqref{eq:specialsymm} for $\rho$ would lead to a cpsdt factorization with real psd matrices of $M$, which is also a cpsd factorization. Such a factorization does not always exists, since being psd is an additional necessary condition (but even this is not sufficient for $d\geq 5$).

(iii) follows from \cref{thm:corresp} and \cref{prop:exspecial}.
\demo
%\end{proof}

This situation is summarised in \cref{tab:symmti}. 
We refer the reader to \cite{De19d} for a more thorough study of the existence of this type of decompositions.

\begin{table}[h]
\begin{tabular}{l|c|c} 
& Of type & Does it always exist? \\
\hline
\multirow{2}{*}{Operator Schmidt decomp.}
& \eqref{eq:specialsymm} & Yes \\
&  \eqref{eq:specialti}  & Yes  \\
\hline
\multirow{2}{*}{Separable decomp.}
&  \eqref{eq:specialsymm}  & No\\
&  \eqref{eq:specialti} &Yes  \\
\hline
\multirow{2}{*}{Purification }
&   \eqref{eq:specialsymm} & Yes \\
&   \eqref{eq:specialti} & Yes \\
\end{tabular}
\caption{Decompositions of bipartite psd matrices, in the symmetric case (of type \eqref{eq:specialsymm})
and in the t.i.\ case (of type \eqref{eq:specialti}).
}
\label{tab:symmti}
\end{table}

%%===============================
\begin{remark}[Decompositions for completely positive maps]
Any bipartite psd matrix $\rho\in \mc{M}_{d_1} \otimes  \mc{M}_{d_2}$ is associated to a completely positive map $\mc{E}:  \mc{M}_{d_2} \to  \mc{M}_{d_1}$ via the the Choi--Jamio\l kowski isomorphism. 
For the particular case of $\sigma$ [Eq.\ \eqref{eq:cb}], the associated completely positive map is 
\be
 \mc{E}(X)= \sum_{i=1}^{d_1} \sum_{j=1}^{d_2}m_{ij} |i\ra \la j|X|j\ra \la i |. 
\ee
In words, $\mc{E}$ projects a matrix $X\in \mc{M}_{d_2}$ to its diagonal elements, and transforms it into another diagonal matrix, thus in fact $\mc{E}:\mc{M}_{d_2} \to \mc{D}_{d_1}$ where $\mc{D}_{d}$ is the set of diagonal matrices of size $d\times d$. If a diagonal matrix is considered as a column vector, this is just matrix multiplication with $M$. That $\mc{E}$ is completely positive implies that, when $\mc{E}$ acts on diagonal matrices, it is a nonnegative map (i.e.\ it maps nonnegative diagonal matrices to nonnegative diagonal matrices). 
Now, the factorisations of $\sigma$  can be interpreted as the following factorisations of $\mc{E}$: 
\begin{itemize}
\item[(i)] $\textrm{rank}(M)$ is the minimal $r$ such that 
$\mc{E}:\mc{D}_{d_2}\to \mc{D}_{r} \to \mc{D}_{d_1}$. 

\item[(ii)]
$\textrm{rank}_+(M)$ is the minimal  $r$ so that 
$\mc{E}:\mc{D}_{d_2}\stackrel{\mc{T}_2}{\to} \mc{D}_{r} \stackrel{\mc{T}_1}{\to} \mc{D}_{d_1}$,
and each $\mc{T}_j$ is a nonnegative map.

\item[(iii)]
 $ \textrm{psd-rank}^\C  (M)$ is the  minimal $r$
such that 
$\mc{E}:\mc{D}_{d_2}\stackrel{\mc{T}_2}{\to} \mc{M}_{r} \stackrel{\mc{T}_1}{\to} \mc{D}_{d_1}$,
and each $\mc{T}_j$ is a completely positive map.

\item[(iv)]
$\symmrank(M)$ is the minimal $r$ so that
$\mc{E}:\mc{D}_{d}\stackrel{\mc{T}}{\to} \mc{D}_{r} \stackrel{\mc{T}'}{\to} \mc{D}_{d}$, 
where $\mc{T}'$ denotes the dual of the map\footnote{The dual is defined by $\tr(Y\mc{T}(X)) = \tr(\mc{T}'(Y)X)$.}.
\item[(v)]
$\textrm{cp-rank}(M)$ is the minimal $r$ so that 
 $\mc{E}:\mc{D}_{d}\stackrel{\mc{T}}{\to} \mc{D}_{r} \stackrel{\mc{T}'}{\to} \mc{D}_{d}$,
and $\mc{T}$ is a nonnegative map.

 \item[(vi)]  $\textrm{cpsdt-rank}^\C (M)$ is the minimal $r$ such that  
$\mc{E}:\mc{D}_{d}\stackrel{\mc{T}}{\to} \mc{M}_{r} \stackrel{\mc{T}'\circ\ \vartheta}{\to} \mc{D}_{d}$,
and  $\mc{T}$ is a completely positive map ($\vartheta$ denotes  the transposition map).
\end{itemize}
Finally note that  the  $\textrm{cpsd-rank}^\C (M)$ is the minimal $r$ such that 
 $$\mc{E}:\mc{D}_{d}\stackrel{\mc{T}}{\to} \mc{M}_{r} \stackrel{\mc{T}'}{\to} \mc{D}_{d}$$
and  $\mc{T}$ is a completely positive map. 
\demo\end{remark}

\begin{remark}{\rm \bf (Other decompositions of nonnegative matrices in quantum information)}
\label{rem:otherconnections}
Decompositions of nonnegative matrices appear in several contexts in quantum information. In the context of games or correlations, 
the nonnegative matrix is a matrix of conditional probabilities $\{p(a,b|x,y)\}_{a,b,x,y}$,  
and the different decompositions and ranks correspond to different strategies (e.g.\ classical versus quantum) of realising these probabilities \cite{Si17}. In that context, the normalisation conditions on the probabilities are important. 
In particular, in Ref.\ \cite{Si17} the authors show that the sets of classical, quantum, no-signaling and unrestricted correlations can be expressed as projections of affine sections of the completely positive cone, the completely positive semidefinite cone, 
the non-signaling cone or the nonnegative cone, respectively. 
\demo\end{remark}

\section{Characterisation of decompositions of psd matrices}
\label{sec:bounds} 

Here we characterise the three decompositions of psd matrices presented in \cref{sec:pos} by proving relations among their ranks, namely $\osr$,  $\seprank$  and $\purirank$. 
We will review some known results (citing the corresponding source) and also provide some new relations (without a citation).  
Many of the results are inspired by the correspondence of \cref{thm:corresp}: 
in some cases they are immediate implications of the corresponding results for nonnegative matrices, 
and in other cases they are generalisations thereof. 
A few results have been derived independently of these connections.

This section is structured as follows: 
we characterise product psd matrices (\cref{ssec:product}),
separable psd matrices (\cref{ssec:sep}),
 pure psd matrices (\cref{ssec:pure}),
and general psd matrices (\cref{ssec:general}). 
Recall that they are all defined in \cref{def:basic}. 
We will then give further upper bounds on the purification rank (\cref{ssec:ubpuri}),
and lower bounds based on the entanglement of purification (\cref{ssec:ep}).

%%------------------------
\subsection{Product psd matrices}
\label{ssec:product}
We start by characterising product psd matrices. The following can be seen as a generalization  of \cite[Proposition 2.8]{Fa14}, by using the correspondence of \cref{thm:corresp}. 

\begin{proposition}[Characterisation of product psd matrices] \label{pro:product}
Let  
$0\leqslant \rho \in \mc{H}^{[1]}\otimes \cdots \otimes  \mc{H}^{[n]}$. 
The following are equivalent:
\begin{itemize}
\item[(i)]$\osr(\rho) = 1 $
\item[(ii)]$\purirank(\rho) = 1$
\item[(iii)]$\seprank(\rho)=1$ 
\end{itemize}
\end{proposition}

%\begin{proof} 
\emph{Proof}.
(i) $\iff$ (iii). 
Let the operator Schmidt decomposition be  $\rho = A^{[1]}\otimes A^{[2]} \otimes \ldots \otimes A^{[n]} $, so that 
$\osr(\rho)=1$. 
 Since $\rho\geqslant 0$, each $A^{[i]}$ is either positive semidefinite or negative semidefinite, and the number of negative semidefinite matrices is even. We can redefine the negative semidefinite matrices as minus themselves without changing $\rho$, and thus obtain that each $A^{[i]}$ is psd. This form is already a separable decomposition with $\seprank(\rho)=1$. The converse direction is immediate. 
 
 (ii) $\iff$ (iii). 
 Consider $\rho = A^{[1]}\otimes \cdots \otimes A^{[n]} $ with $A^{[l]}\geqslant 0$, so that $\seprank(\rho)=1$. 
 To see that $\purirank(\rho) = 1$, write $A^{[l]} = C^{[l]}C^{[l]\dagger}$, and define $L = C^{[1]} \otimes \cdots \otimes C^{[n]}$. This satisfies that $\osr(L)=1$ and $LL^\dagger = \rho$, and therefore $\purirank(\rho) = 1$. 
  The converse direction is immediate. 
\demo
%\end{proof}

\begin{example}
$\rho = \frac{I}{2} \otimes |0\ra\la 0| \otimes |+\ra\la +|$ is a product  state (where $|+\ra = (|0\ra+|1\ra)/\sqrt{2}$ and $I = |0\ra\la 0|+|1\ra\la 1|$). From this expression, it has $\osr(\rho)=1$. Since every term is psd, this expression is also a separable decomposition, and thus $\seprank(\rho)=1$. The matrix $L = \frac{I}{\sqrt{2}} \otimes |0\ra\la 0| \otimes |+\ra\la +| $ is such that $LL^\dagger = \rho$ and has $\osr(L)=1$, and thus $\purirank(\rho)=1$.
\end{example}
%%------------------------
\subsection{Separable psd matrices}
\label{ssec:sep}

We now characterise separable psd matrices. 

\begin{proposition}{\rm \bf (Characterisation of separable psd matrices)}
\label{pro:sep} 
Let  $0\leqslant \rho \in \mc{H}^{[1]}\otimes \cdots \otimes  \mc{H}^{[n]}$ be separable (\cref{def:sep}). 
Then 
\begin{itemize}
\item[(i)]$\osr(\rho)\leq \seprank(\rho)$. 
\item[(ii)]$\purirank(\rho)\leq \seprank(\rho)$, but $\seprank(\rho)$ cannot be upper bounded by a function of $\purirank(\rho)$ only. 
 \end{itemize}
\end{proposition}

%\begin{proof}
\emph{Proof}.
(i) follows from the fact that the separable decomposition is a special case of the MPDO form.

(ii) The inequality follows from the fact that the separable decomposition (\cref{def:sepdecomp}) is a special case of the local purification form [Eq.\ \eqref{eq:lp}], in which $\alpha_i=\beta_i$ for $i=1,\ldots, n-1$. We show it explicitly for the bipartite case. So consider the separable decomposition 
$$
\rho = \sum_{\alpha=1}^s \sigma_\alpha\otimes \tau_\alpha
$$
where $s=\seprank(\rho)$ and $\sigma_\alpha\geqslant 0$ and  $\tau_\alpha\geqslant 0$. 
We write $\sigma_\alpha = A_\alpha A_\alpha^\dagger$ and  $\tau_\alpha = B_\alpha B_\alpha^\dagger$, and define the matrix 
$$
L = \sum_{\alpha=1}^s (A_\alpha \otimes \la \alpha|) \otimes B_\alpha
$$
This verifies that $LL^\dagger = \rho$, and thus $L$ is a purification of $\rho$ with $\osr(L)\leq s$. Thus $\purirank(\rho)\leq \seprank(\rho)$. 

The separation between $\seprank(\rho)$ and  $\purirank(\rho)$ follows from \cref{thm:corresp}, and the fact that 
there is a separation between $\rank_+$ and $\psdrank$ \cite{Go12}. 
That is,  $\psdrank$ cannot be upper bounded by a function of $\rank_+$  only. 
\demo
%\end{proof}

In the special case of $n=2$, if $\osr(\rho)=2$, then $\rho$ is separable, and we know exactly the values of the three ranks: 

\begin{proposition}[\cite{De19c}]\label{pro:osr2}
Let  $0\leqslant \rho \in \mc{H}^{[1]}\otimes  \mc{H}^{[2]}$. 
If $\osr(\rho)=2$, then $\rho$ is separable with $\seprank(\rho)=2$. Thus, $\purirank(\rho)=2$ as well. 
\end{proposition}

\begin{example}Consider the bipartite matrix  
$$
\rho = I\otimes I + \sigma_x \otimes \sigma_x.
$$
This has $\osr(\rho)=2$ and is thus separable. In this case the separable decomposition can be found by inspection, namely
\be
\rho = \frac{1}{2}(|+,+\ra \la +,+| +|-,-\ra \la -,-|  ),
\label{eq:rhosep+}
\ee 
where $|\pm \ra = (|0\ra\pm|1\ra)/\sqrt{2}$,  which indeed has $\seprank(\rho)=2$. 

\end{example}

%%================================
\subsection{Pure psd matrices}
\label{ssec:pure}
We now characterise pure psd matrices, that is, $\rho$ with $\rank(\rho)=1$. 
In this case $\rho = LL^\dagger$ where $L$ is a column vector.

\begin{proposition}[Characterisation of pure psd matrices]
\label{pro:pure}
$0\leqslant \rho \in \mc{H}^{[1]}\otimes \cdots \otimes  \mc{H}^{[n]}$ be a pure psd matrix, i.e.\ $\rho = L L^\dagger$ where $L$ is a column vector. Then 
\begin{enumerate}
\item[(i)] $\osr(\rho) = \osr(L)^2$.
\item[(ii)]$\purirank(\rho) = \osr(L)$. 
\end{enumerate}
\end{proposition}

See \cref{rem:mpo} concerning the use of operator Schmidt rank for vectors. 
 
%\begin{proof}
\emph{Proof}.
We prove it for the bipartite case for ease of notation. In addition, we write $|L\ra$ instead of $L$ in order to emphasise that $L$ is a column vector. 
The same will be true of other matrices appearing in this proof. 
(i) 
Consider the operator Schmidt decomposition $|L\ra = \sum_{\alpha=1}^s |A_\alpha\ra \otimes |B_\alpha\ra $ where $s=\osr(L)$, and thus 
$\rho = \sum_{\alpha,\beta=1}^s |A_\alpha\ra \la A_\beta| \otimes |B_\alpha\ra \la B_\beta|$. 
This shows that $\osr(\rho)\leq s^2$. 
Since $\{|A_\alpha\ra \}_{\alpha}$ and $\{|B_\alpha\ra \}_{\alpha}$ are linearly independent, so are $\{|A_\alpha\ra \la A_{\beta}|\}_{\alpha,\beta}$ and 
$\{|B_\alpha\ra \la B_{\beta}|\}_{\alpha,\beta}$, and thus the dimension of their span is $s^2$. Thus $\osr(\rho)=s^2$. 

(ii) That $\purirank(\rho) \leq \osr(L)$ is clear, since $L $ provides a purification where the auxiliary system has dimension one. Since $\{|A_\alpha\ra \la A_{\beta}|\}_{\alpha,\beta}$ and $\{|B_\alpha\ra \la B_{\beta}|\}_{\alpha,\beta}$ are linearly independent, there cannot exist another purification $L'$ with $\osr(L')<\osr(L)$. 
\demo
%\end{proof}

Thus, for a pure psd matrix $\rho = LL^\dagger$ (with $L$ a column vector) we have that
$$
\osr(\rho) = \osr(L)^2 =\purirank(\rho)^2 .
$$

\begin{example}
Consider the pure state $\rho = |\Phi^+\ra \la \Phi^+|$, where $L= |\Phi^+\ra = \frac{1}{\sqrt{2}}(|00\ra + |11\ra)$.  This is a purification with $\osr(L) =\purirank(\rho) =2 $. 
It is immediate to see that  $\osr(\rho) = 4$. 
\end{example}

%%===========================
\subsection{General psd matrices}
\label{ssec:general}

We now consider the general case, that is, where $\rho$ need not be product, separable or pure. 
First we  will bound the various ranks in terms of the physical dimension of $\rho$ (\cref{pro:physdim}),
 and then characterise the relation between $\osr$ and $\purirank $ (\cref{pro:osrpurirank} and \cref{pro:separationosrpuri}).

\begin{proposition}{\rm \bf (Bounds in terms of the physical dimension)} \label{pro:physdim}
Let $0\leqslant \rho \in \mc{H}^{[1]}\otimes \cdots \otimes\mc{H}^{[n]}$, where each $\mc{H}^{[j]}$ is given by $\mc{M}_d$. The following relations hold:
\bi
\im[(i)] 
$\osr(\rho) \leq d^{2\lfloor n/2\rfloor}$.
\im[(ii)]  
$\purirank(\rho)\leq d^{2\lfloor n/2\rfloor}$. 
\im[(iii)] If $\rho$ is separable, then 
$\seprank(\rho)\leq d^{2n}$. 
\ei
\end{proposition}

The bounds can be extended in a straightforward way to the case that each local Hilbert space has a different dimension, i.e.\ where $\mc{H}^{[j]}$ is given by $\mc{M}_{d_j}$. 

%\begin{proof}
\emph{Proof}.
(i)  This follows from the construction of \cref{ssec:mpdo}, since there we obtained that at bipartition  $[1,\ldots, l] |[l+1, \ldots, n]$, 
$$
D_{l}\leq \min\{d^2 D_{l-1},d^{2l},d^{2(n-l)} \}.
$$
(ii) By \cref{pro:purisqrt} we have that $\purirank(\rho)\leq \osr(\sqrt{\rho})$. 
Since $\sqrt{\rho}\in \mc{M}_{d^n}$,  (i) implies that  $\osr(\sqrt{\rho}) \leq d^{2\lceil n/2\rceil}$. 
(iii) 
We have 
$\rho\in (\Her_d)^{\otimes n} = \Her_{d^n}$, which is a real vector space of dimension $d^{2n}$. 
By Caratheodory's Theorem, $\rho$ can be written as a sum of at most $d^{2n}$ elements of product psd matrices. 
 This is an upper bound on the ``tensor rank version" of the separable rank, which itself upper bounds the  separable rank, and thus proves the claim. 
\demo
%\end{proof}

We now turn to the relation between $\osr(\rho)$ and $\purirank(\rho)$. 

\begin{proposition} \label{pro:osrpurirank} 
Let $0\leqslant \rho \in \mc{H}^{[1]}\otimes \cdots \otimes  \mc{H}^{[n]}$. Then 
$\osr(\rho)\leq \purirank(\rho)^2$, and the bound is tight for pure psd matrices.  
\end{proposition}

%\begin{proof}
\emph{Proof.}
For pure psd matrices we have that $\osr(\rho)= \purirank(\rho)^2$ by \cref{pro:pure}. 
For a general $\rho$, consider an optimal purification $L$, i.e.\ such that $L L^\dagger =\rho $ and  $\osr(L) =\purirank(\rho)$.
Then 
\begin{equation*}
\osr(\rho) = \osr(LL^\dagger)\leq  \osr(L)\osr(L^\dagger) =\osr(L)^2 =\purirank(\rho)^2.%\qedhere
\end{equation*}
\demo
%\end{proof}

\begin{proposition}{\rm \bf (Separation result for entangled psd matrices)}\cite{De13c,Go12} \label{pro:separationosrpuri}
 $\purirank$ cannot be upper bounded  by a function of $\osr$ only. 
 In particular, there is a sequence of psd matrices  diagonal in the computational basis $(\rho_n \in \mc{M}_{2}^{\otimes (2n)} )_n$, with $\osr(\rho_n)=3$ for all $n$, and $\purirank(\rho_n)>\Omega(\log n)$.
 \label{thm:separation}
\end{proposition}

%\begin{proof}
\emph{Proof}.
Ref.\ \cite{Go12} shows that $\rank_{\textrm{psd}}(M)$ cannot be upper bounded by  $\rank(M)$ only. In particular, Ref.\ \cite{Go12} shows that if $S_t$ is the slack matrix of  the regular $t$-gon, then $\rank(S_t) =3$ (for $t\geq 3$) and $\psdrank(S_t) > \Omega(\log  t)$.

Using the correspondence of \cref{thm:corresp}, this implies that  there is a sequence of bipartite psd matrices diagonal in the computational basis whose purification rank cannot be upper bounded by its operator Schmidt rank. This counterexample is extended to the multipartite case in \cite{De13c}. 
\demo
%\end{proof}

%%=========================
\subsection{Upper bounds on the purification rank}
\label{ssec:ubpuri}
Given the separation result of \cref{pro:separationosrpuri}, it is interesting to study how the purification rank can be upper bounded. We now provide several such upper bounds. 

We start by relating the purification rank with the quantum square root rank (\cref{def:qsqrtrank}). This result is the analogue of \cite[Theorem 2.9 (v)]{Fa14} for psd matrices. 

\begin{proposition} \label{pro:purisqrt} 
Let $0\leqslant \rho\in \mc{H}^{[1]}\otimes \cdots \otimes  \mc{H}^{[n]}$. Then $\purirank(\rho)\leq \qsqrtrank(\rho)$, but 
$\qsqrtrank(\rho)$ cannot be upper bounded by a function of $\purirank(\rho)$ only. 
\end{proposition}

%\begin{proof}
\emph{Proof.}
Concerning  the first statement, simply note that in $\qsqrtrank(\rho)$ one minimises over the square roots of $\rho$, whereas in $\purirank(\rho)$ (\cref{def:localpurif}) one minimizes over matrices $L$ such that $LL^\dagger =\rho$, which in particular includes the square roots of $\rho$. 
The second statement follows from the fact that the square root rank cannot be upper bounded by a function of the psd rank \cite{Fa14}, and from \cref{thm:corresp}. 
\demo
%\end{proof}

The following results upper bound  the purification rank by a function of the operator Schmidt rank \emph{and} the physical dimension $d^n$  \cite{De13c}. Here we prove them in the  language of this paper.

\begin{proposition}{\rm \bf (Bounds of $\purirank$ in terms of the physical dimension and $\osr$)}\cite{De13c}
Let $0\leqslant \rho \in \mc{H}^{[1]}\otimes \cdots \otimes \mc{H}^{[n]}$ where each $\mc{H}^{[j]}$ is given by $\mc{M}_d$. The following holds:
\bi
\im[(i)] \label{prop:sos}
Let $m$ be the number of different eigenvalues of $\rho$ including 0, so that 
$m \leq \min\{\rank(\rho)+1,d^n\}$. 
Then $$
\purirank(\rho)\leq \qsqrtrank(\rho)\leq 
\frac{\osr(\rho)^m-1}{\osr(\rho)-1}. 
$$
\im[(ii)] $\purirank(\rho)\leq \osr(\rho) (\rank(\rho))^2$.
\ei
\end{proposition}

%\begin{proof}
\emph{Proof.}
(i)  We clearly have 
$$
\purirank(\rho) \leq \qsqrtrank(\rho) \leq \osr(\sqrt{\rho})
$$
where $\sqrt{\rho}\geqslant 0$, and where the first inequality follows from \cref{pro:purisqrt}. 
There is a polynomial $p$ of degree $m-1$, such that 
$\sqrt{\rho} = p(\rho). $
We thus have that
$$
\osr(\sqrt{\rho}) = \osr(p_m(\rho))  \leq \sum_{l=0}^{m-1} \osr(\rho^l)  \leq \sum_{l=0}^{m-1} \osr(\rho)^l = \frac{\osr(\rho)^m-1}{\osr(\rho)-1}. 
$$

(ii) 
Consider the spectral decomposition of $\rho = \sum_{j=1}^s\lambda_j |\psi_j\ra\la\psi_j|$, with $s =\rank(\rho)$. 
The ``standard'' choice of $L$ is $L = \sum_{j=1}^s \sqrt{\lambda_j} |\psi_j\ra \la j|$, as this clearly satisfies $\rho=LL^\dagger$. 
For this $L$ we have that $\purirank(\rho)\leq \osr(L)$, and in the following we will upper bound the latter. 

Consider a product state $|p_\alpha\ra$, so that by definition $\osr(|p_\alpha\ra) =1$. 
Define 
$$
|\chi_\alpha\ra :=\rho |p_\alpha\ra = LL^\dagger |p_\alpha\ra. 
$$
By construction we have that $\osr(|\chi_\alpha\ra)\leq \osr(\rho)$.

Now choose $s$ product states $\{|p_\alpha\ra\}$ 
so that $\{L^\dagger |p_\alpha\ra\}_{\alpha=1}^s$ are linearly independent. 
Thus we can express $|\psi_j\ra = \sum_{\alpha=1}^s c_{j,\alpha} |\chi_\alpha\ra$ for some coefficients $c_{j,\alpha}$, 
so that $\osr(|\psi_j\ra)\leq  \osr(\rho) \: s$, and finally
$$
\osr(L)\leq s \max_j [ \osr( |\psi_j\ra) \osr( \la j|  ) ]\leq \osr(\rho) s^2,
$$ 
the desired result.
\demo
%\end{proof}

%%===================
\subsection{Lower bounds based on the entanglement of purification}
\label{ssec:ep}
We now show that, in the bipartite case, we can lower bound the purification rank using the entanglement of purification \cite{Te02}. The entanglement of purification is a measure of classical and quantum correlations, which, for a bipartite state $\rho_{AB}$ is defined as 
\be
E_p(\rho_{AB}) = \min_{\psi} \{E(|\psi\ra_{A,A',B,B'} ))| \tr_{A'B'} |\psi\ra\la \psi| = \rho\} , 
\nn
\ee
where the entropy of entanglement is defined as 
$$
E(|\psi\ra_{A,A',B,B'} ) = S_1(\rho_{AA'}), 
$$ 
where $\rho_{AA'} = \tr_{BB'} |\psi\ra_{AA'BB'}\la \psi| $ and 
where $S_1$ is the von Neumann entropy, $S_1(\rho) =-\tr(\rho\log\rho)$. 

\begin{proposition}{\rm \bf (Bound  in terms of the entanglement purification)} Let $\rho_{AB}$ be a bipartite state. Then 
$E_p(\rho_{AB}) \leq \log (\purirank(\rho_{AB})$). 
\end{proposition}

%\begin{proof}
\emph{Proof.}
In the bipartite case the purification rank is defined as 
$$
\purirank(\rho_{AB}) = \min_{\psi} \{ \sr(|\psi\ra ) | \tr_{A'B'} |\psi\ra\la \psi| = \rho\},
$$
where the $\sr(|\psi\ra )$ denotes the Schmidt rank of $|\psi\ra$  across bipartition $AA'|BB'$.
Noting that $\sr(|\psi\ra)  = \rank(\rho_{AA'})$ and recalling that the Renyi entropy with parameter $\alpha=0$ is defined as is $S_0(\rho) = \log\rank (\rho) $ we have that 
$$
\log ( \purirank(\rho_{AB})) = \min_{\psi} \{ S_0(\rho_{AA'}) | \tr_{A'B'} |\psi\ra\la \psi| = \rho\}.
$$
Let $|\phi\ra$ denote the state that achieves the minimisation in the previous equation. Then we have that
\be
\nonumber\log ( \purirank(\rho_{AB})) &=& S_0(\tr_{BB'}(|\phi\ra\la \phi|))\\
\nonumber&\geq & S_1(\tr_{BB'}(|\phi\ra\la \phi|))\\
\nonumber&\geq & \min_{\psi} \{S_1(\tr_{BB'}(|\psi\ra\la \psi|) | \tr_{A'B'}|\psi\ra\la\psi|=\rho \} \\
\nonumber &=& E_p(\rho_{AB}) ,
\ee
where we have used that $S_0(\rho)\geq S_1(\rho)$. 
\demo
%\end{proof}

Thus any lower bound of the entanglement of purification also lower bounds the logarithm of the purification rank.

%%===========================
\section{Characterisation of decompositions of t.i.\ psd matrices}
\label{sec:tibounds}

In this section we characterise the decompositions of t.i.\ psd matrices (\cref{def:ti}).
In the bipartite case we will focus exclusively on decompositions of type \eqref{eq:specialti}. 
We first characterise t.i.\ product psd matrices, 
 t.i.\ separable psd matrices 
 and  general t.i.\  psd matrices (\cref{ssec:tigeneral}), and 
then give relations with their non-t.i.\ counterparts (\cref{ssec:nti}).
As in the previous section, we will review some results (citing the source), and mostly prove new relations (without a citation).

%%========
\subsection{T.i.\ psd matrices}
\label{ssec:tigeneral}

First, for t.i.\ product psd matrices the analogue of \cref{pro:product} is true:

\begin{proposition}[Characterisation of t.i.\ product states]\label{pro:tiproduct}
Let $0\leqslant \rho \in (\mc{H}_{\mathrm{l}})^{\otimes n}$ be  t.i. 
The following are equivalent:
\begin{itemize}
\item[(i)] $\tiosr(\rho) = 1$
\item[(ii)] $\tipurirank(\rho) = 1 $
\item[(iii)] $\tiseprank(\rho)=1$
\end{itemize}
\end{proposition}

%\begin{proof} 
\emph{Proof.}
Analogous to the proof of \cref{pro:product}. 
\demo
%\end{proof}

For t.i.\ separable psd matrices (\cref{def:tisep}) the analogue of \cref{pro:sep} is true: 

\begin{proposition}{\rm \bf (Relation between of $\tipurirank$ and $\tiseprank$)}
Let $0\leqslant \rho \in (\mc{H}_{\mathrm{l}})^{\otimes n}$ be a t.i.\ separable psd matrix (\cref{def:tisep}).
 Then $\tipurirank(\rho)\leq\tiseprank(\rho)$.  
\end{proposition}

%\begin{proof} 
\emph{Proof.}
Analogous to the proof of \cref{pro:sep}. 
\demo
%\end{proof}

We now characterise t.i.\ psd matrices $\rho$ which are not necessarily product or separable. 
Note again that any such matrix has a t.i.\ local purification.

\begin{proposition}
For $\rho\in (\mc{H}_\mathrm{l})^{\otimes n}$ and any t.i.\ $L$ with $LL^\dagger=\rho$ we have $$
\tipurirank(\rho)\leq \tiosr(L)\leq n\cdot  \osr(L). 
$$
\end{proposition}

%\begin{proof} 
\emph{Proof.}
The first inequality is trivial, the second follows from the construction of \cref{rem:ti} to $L$.
\demo
%\end{proof}

We now give the analogue of \cref{pro:osrpurirank}.

\begin{proposition} \label{pro:tiosrpurirank}
Let $0\leqslant\rho\in (\mc{H}_\mathrm{l})^{\otimes n}$ be t.i. Then 
$\tiosr(\rho)\leq \tipurirank(\rho)^2$. 
\end{proposition}

%\begin{proof} 
\emph{Proof.}
Analogous to the proof of \cref{pro:osrpurirank}. 
\demo
%\end{proof}

%%--------=========--------------=====================
The previous proposition relates $\tiosr$ with $\tipurirank$ for a fixed system size $n$. 
It is also interesting to study whether  $\tiosr$ with $\tipurirank$ can be related in a way that is independent of $n$, i.e.\ uniform in $n$. 
In the following, we review a negative result in this direction  \cite{De15}. 

In order to state our result we need a few definitions first.
Given a tensor $A=\{A_{\alpha,\beta}\in \mc{M}_{d}\}_{\alpha,\beta=1}^D$,  define 
$$
\rho_n(A) := \sum_{\alpha_1,\ldots, \alpha_n=1}^D 
A_{\alpha_1,\alpha_2} \otimes A_{\alpha_2,\alpha_3} \otimes \cdots \otimes A_{\alpha_n,\alpha_1}  ,
$$

\begin{theorem}[\cite{De15}]
\label{thm:undec} 
Not for every tensor $A$ such that $\rho_n(A)\geqslant  0$ for all $n\geq 1$, there is a tensor $B$ such that
$$
\rho_n(A) \propto \rho_n(B)^\dagger \rho_n(B) \qquad \forall n.
$$
\end{theorem}

Here $\propto$ means ``proportional to'', that is, there is a positive constant $c_n$ so that $\rho_n(A)=c_n \rho_n(B)^\dagger \rho_n(B)$.
The theorem says that even if there is a tensor $A$ that gives rise a family of psd matrices $\rho_n(A)$ for all $n$, there may not exist another $B$ (of any finite size) which provides a t.i.\ local purification of $\rho_n(A)$ \emph{which is valid for all system sizes}. This is true even if we allow for a different proportionality constant $c_n$ for each $n$. 
In other words, the theorem says that there are psd matrices that admit a t.i.\ MPDO form independent of the system size, but no t.i.\ local purification form independent of the system size. 
This is true even if $\rho_n(A)$ is diagonal in the computational basis, i.e.\ if $A^{i,j}_{\alpha,\alpha'}\propto \delta(i,j)$  (where $\delta(i,j)$ is the Kronecker delta) and $(d,D)\geq (7,7)$ \cite{De15}.

The idea of the proof is the following. 
Ref.\ \cite{De15} shows that the assumption ``$\rho_n(A) \geqslant 0$ for all $n$"  is in fact undecidable, i.e.\ given a  tensor $A$, there is no algorithm that decides whether $\rho_n(A)\geqslant 0$ for all $n$. 
Now assume that there is a $B$ such that $\rho_n(A) \propto \rho_n(B)\rho_n(B)^\dagger$ for all $n$. 
Ref.\ \cite{De15} provides an algorithm to find this $B$. This allows to verify that $\rho_n(A)\geqslant 0$, and thus to solve an undecidable problem. 
Thus, this $B$ cannot exist in general.

Note that \cref{thm:undec} is proven without any reference to the corresponding decompositions of nonnegative matrices (correspondences of \cref{thm:corresp}), albeit they may be related.

Finally we remark  that similar problems but with open boundary conditions have been shown to be undecidable \cite{Kl14}, and NP-complete for finitely many $n$ \cite{Kl14}, and similar results have been obtained for  two dimensional structures of tensor networks \cite{Sc18}.

%%========
\subsection{Relation to the non-t.i.\ counterparts}
\label{ssec:nti}

We now give relations between the t.i.\ ranks and their non-t.i.\ counterparts. 
We start with the relation between $\osr$ and $\tiosr$.

\begin{proposition}[Relation between $\osr$ and $\tiosr$]\label{pro:tintiosr} 
Let $0\leqslant \rho \in (\mc{H}_\mathrm{l})^{\otimes n}$ be t.i. Then 
$$
\osr (\rho ) \leq \tiosr (\rho )\leq  n \: \osr(\rho) .
$$
\end{proposition}

%\begin{proof}
\emph{Proof.}
The first inequality follows from the fact that the t.i.\ MPDO is a special case of the MPDO form in which the local tensors are independent of the site. 
The second inequality  follows from the construction of \cref{rem:ti}. 
\demo
%\end{proof}

%%----------------------------

We now turn to the relation between $\purirank$ and $\tipurirank$.

\begin{proposition}{\rm \bf (Relation between $\purirank$ and $\tipurirank$)}
\label{pro:puriranktinonti} 
Let $0\leqslant \rho \in (\mc{H}_{\mathrm{l}})^{\otimes n}$ be  t.i. 
Then 
$$\purirank(\rho)\leq \tipurirank(\rho).$$
\end{proposition}

%\begin{proof}
\emph{Proof.}
Simply note that the t.i.\ local purification form  is a special case of the local purification in which the local tensors are independent of the site. 
\demo
%\end{proof}

Note that we cannot use the construction of \cref{rem:ti} to 
upper bound $\tipurirank$ by $\purirank$, since $L$ need not be t.i., as discussed in \cref{ssec:tigeneral}.

%%----------------------------
We now turn to the $\tiseprank$ and its relation to $\seprank$.

\begin{proposition}{\rm \bf (Relation between $\seprank$ and $\tiseprank$)} \label{pro:sepranktiseprank} 
Let $0\leqslant \rho\in (\mc{H}_\mathrm{l})^{\otimes n}$ be a t.i.\ separable psd matrix (\cref{def:tisep}). Then 
$$
\seprank(\rho)\leq \tiseprank(\rho)\leq n \: \seprank(\rho)
$$
and the first inequality is tight. 
\end{proposition}

%\begin{proof}
\emph{Proof.}
The first inequality is obvious, since the t.i.\ separable decomposition is a special case of the separable decomposition. 
To see that it is tight, simply note that for a t.i.\ product state we have 
$\tiseprank(\rho)=1$ by \cref{pro:tiproduct} and thus also $\seprank(\rho)=1$.  
The second inequality follows from the construction of \cref{lem:tiseptidecomp}. 
\demo
%\end{proof}

We do not know whether the second inequality in \cref{pro:sepranktiseprank} is tight.
 \cref{ex:modifiedW} provides a state for which 
$$
\tiseprank(\rho)\geq \frac{\sqrt{n}}{2} \: \seprank(\rho).
$$ 

%%----------------------------
The previous three results (\cref{pro:tintiosr}, \cref{pro:puriranktinonti} and \cref{pro:sepranktiseprank}) imply that if we have a t.i.\ psd matrix and we  enforce translational invariance in its representation, this will generally increase the cost by  an amount that depends on the system size. This holds for the three t.i.\  representations considered here: the t.i.\ MPDO form, the t.i.\ local purification form (if it exists), and the t.i.\ separable form (if $\rho$ is separable).

%%=================
%%===========================
\section{Conclusions and Outlook}
\label{sec:concl}

In this paper we have studied several natural decompositions of positive semidefinite matrices $\rho$ with a one-dimensional structure. 
We have considered and characterised the MPDO form, the separable form and the local purification, 
as well as  their corresponding ranks ($\osr, \seprank,\purirank$).
We have also defined and characterised their translationally invariant (t.i.) analogues: 
the t.i.\ MPDO form, the t.i.\ separable form and the t.i.\ local purification, as well as their corresponding ranks ($\tiosr, \tiseprank, \tipurirank$). In the bipartite t.i.\ case, we have considered the symmetric versions of these decompositions. 

For bipartite states which are diagonal in the computational basis, 
we have presented a correspondence between these decompositions and factorisations of nonnegative matrices (\cref{thm:corresp}). 
We have leveraged this correspondence to derive several  bounds and relations between the different ranks, presented in \cref{sec:bounds} and \cref{sec:tibounds}. 

Note that Ref.\ \cite{Gl19} proves some of the results of \cref{thm:corresp}, with a focus on applications on machine learning (Ref.\ \cite{Gl19} coincidentally appeared on the arxiv on the same day as this paper). 

We remark that the results of this paper have been greatly generalised in \cite{De19d}, where  
tensor decompositions not only on one-dimensional spatial structures, but on arbitrary simplicial complexes are considered. The translational invariant symmetry considered in this paper is lifted to an invariance given by an arbitrary symmetry group, and the existence and the associated rank of these invariant tensor decompositions is studied. Ref.\  \cite{De19d} also proves a much more general version of \cref{thm:corresp}, and explores some of their consequences regarding separations and generalisations.

Beyond these results, 
some straightforward open questions are the following: 
\begin{itemize}
\item It is known that there is a separation between $\rank$ and $\psdrank$, and between $\rank_+$ and $\psdrank$, which can be represented by the symbol $\ll$ (see, e.g.\ \cite{Fa14}): 
$$
\rank \ll \psdrank \ll \rank_+
$$
Via \cref{thm:corresp}, this immediately implies a separation between $\osr$ and $\purirank$, and between $\purirank$ and $\seprank$, respectively, 
$$
\osr \ll \purirank \ll \seprank .
$$
But can the separations be stronger between $\osr$ and $\purirank$, or between $\purirank$ and $\seprank$, than for their counterparts above?

\item 
\cref{pro:submulti} has simply used the fact that $\psdrank$ is submultiplicative to conclude that  $\purirank$ is too. But can there be stronger differences between 
$\purirank(\rho) \purirank(\rho')$ than for the $\psdrank$?
\end{itemize}

One could also study generalisations of 
the upper bounds for the cp rank of Ref.\ \cite{Be03},
the lower bounds to the nonnegative and cp rank \cite{Fa16}, 
or lower bounds to the cpsd rank \cite{Gr17}. 

Another interesting direction concerns the implications of the computational complexity results of  factorisations of nonnegative matrices \cite{Sh16} for decompositions of quantum states. 
Also, it has been recently discovered that the set of quantum correlations is not closed \cite{Sl17}, from which it follows that the set of cpsd matrices is not closed \cite{Si17}. It would be worth investigating the consequences of that for quantum states. 

A further perspective concerns the study of approximate versions of the various decompositions, where one could investigate whether the separations between the ranks also hold in the approximate case. Preliminary results\cite{De20} suggest that all separations between ranks disappear for approximations in the Schatten $p$-norm with $p>1$.

%\begin{acknowledgments}
%\end{acknowledgments}

\appendix

%%%==============================================
\section{Proof of \cref{thm:corresp}}
\label{app:proofmain}

%\begin{proof}
\emph{Proof of \cref{thm:corresp}}.
We will show  the equivalence of the ranks. The equivalence of the decompositions will be obvious from each of the proofs. 

(i)  Consider a minimal factorisation $M=AB$, i.e.\  $m_{ij} =  \sum_{k=1}^r a_{ik}b_{kj}$,  where $r =\rank(M)$, and where $a_{ik},b_{kj}$ are real numbers. 
Substituting in the expression of $\sigma$ we obtain
$$
\sigma = \sum_{k=1}^r \left(\sum_{i=1}^{d_1} a_{ik}|i\ra\la i|\right) \otimes \left(\sum_{j=1}^{d_2} b_{kj}|j\ra\la j|\right) .
\label{eq:decom}
$$
This shows that $\osr(\sigma)\leq \rank(M)$. To see the opposite direction, 
consider an MPDO decomposition of $\sigma$, 
$$
\sigma = \sum_{k=1}^r A^{[1]}_{k}\otimes A^{[2]}_{k}.
$$ 
Considering the matrix element $ii$ ($jj$) of the first (second) tensor factor, 
 we obtain
$$
m_{ij} = \sum_{k=1}^r (A^{[1]}_{k})_{ii} (A^{[2]}_{k})_{jj}, 
$$
which  shows that $ \rank(M) \leq \osr(\sigma)$.

(ii) This is done precisely as in (i), using that psd matrices have nonnegative diagonal entries and diagonal nonnegative matrices are psd.

(iii) Consider a psd factorisation of $M$, $m_{ij} = \tr(E_iF_j^t)$, where $E_i$ and $F_j$ are psd matrices of size $r$. Write down Gram decompositions $$E_i=\left(a_{ik}^\dagger a_{il}\right)_{k,l} \qquad F_j=\left(b_{jk}^\dagger b_{jl}\right)_{k,l}$$ and define new matrices \begin{align*}L_k^{[1]}&:=\sum_{i=1}^{d_1} |i\rangle\langle i|\otimes a_{ik}^\dagger \in\mathcal M_{d_1,rd_1}(\mathbb C)\\  L_k^{[2]}&:=\sum_{j=1}^{d_2} |j\rangle\langle j|\otimes b_{jk}^\dagger\in\mathcal M_{d_2,rd_2}(\mathbb C)\end{align*} for $k=1,\ldots, r.$  For $$L:=\sum_{k=1}^r L_k^{[1]} \otimes L_k^{[2]}$$ we obtain $LL^\dagger=\sigma.$ This proves ${\rm puri\mbox{-}rank}(\sigma)\leq  \psdrank(M).$

Conversely, let $\sigma=LL^\dagger$ with   $$L=\sum_{k=1}^r L_k^{[1]} \otimes L_k^{[2]}\in\mathcal M_{d_1,d_1'}(\C)\otimes \mathcal M_{d_2,d_2'}(\C)$$ be a purification of $\sigma.$  The matrices \begin{align*}E_i&:=\left(\sum_{s=1}^{d_1'} \left(L^{[1]}_k\right)_{is}\left(L_l^{[1]\dagger}\right)_{si}\right)_{k,l}\in\mathcal M_r(\C)\qquad i=1,\ldots, d_1 \\ F_j&:=\left(\sum_{s=1}^{d_2'} \left(L^{[2]}_k\right)_{js}\left(L_l^{[2]\dagger}\right)_{sj}\right)_{k,l}\in\mathcal M_r(\C)\qquad j=1,\ldots, d_2 \end{align*}
then provide a psd-factorisation with matrices of size $r$ of $M$, as a straightforward computation shows. 
This proves ${\rm puri\mbox{-}rank}(\sigma)\geq  \psdrank(M).$

(iv) is proven exactly as in (i), using $B=A^t$ for one direction and $A_k^{[1]}=A_k^{[2]}$ for the other.
(v) follows from (iv) in the same way as (ii) followed from (i).
(vi) is proven as (iii), but using $F_i=E_i$ for one direction and $L_k^{[1]}=L_k^{[2]}$ for the other.

(vii) We write $\sigma = L^2$, and note that $L$ must be diagonal in the computational basis as well \cite{Hi08}.
Writing $L= \sum_{i,j}n_{ij} |i\ra\la i|\otimes |j\ra\la j|$ and defining  $N=\sum_{i,j} n_{ij}|i\ra\la j|$, the result is immediate. 
\demo
%\end{proof}

%\bibliography{/Users/gemmadelascuevas/Dropbox/Gemma/Special-files/all-my-bibliography.bib}

%%====== BIBLIOGRAPHY =========

\end{document}